\documentclass[aps,prd,twocolumn,showpacs,preprintnumbers,superscriptaddress,nofootinbib,amsmath,amssymb]{revtex4}

\usepackage{graphicx}
\usepackage{dcolumn}
\usepackage{bm}
\usepackage{amsmath}
\usepackage[normalem]{ulem}

\newcommand{\beq}{\begin{eqnarray}}
\newcommand{\eeq}{\end{eqnarray}}

\newcommand{\tr}{\ensuremath{\mathrm{Tr}}}

\newcommand{\ds}{\ensuremath{\displaystyle}}

\def\spose#1{\hbox to 0pt{#1\hss}}
\def\ltapprox{\mathrel{\spose{\lower 3pt\hbox{$\mathchar"218$}}
 \raise 2.0pt\hbox{$\mathchar"13C$}}}

\begin{document}

\title{
Roberge-Weiss endpoint and chiral symmetry restoration in 
$N_f = 2+1$ QCD
}

\author{Claudio Bonati}
\email{claudio.bonati@df.unipi.it}
\affiliation{Universit\`a di Pisa and INFN Sezione di Pisa,\\ 
Largo Pontecorvo 3, I-56127 Pisa, Italy
}
\author{Enrico Calore}
\email{enrico.calore@fe.infn.it}
\affiliation{Universit\`a degli Studi di Ferrara and INFN Sezione di Ferrara,\\
Via Saragat 1, I-44122 Ferrara, Italy
}
\author{Massimo D'Elia}
\email{massimo.delia@unipi.it}
\affiliation{Universit\`a di Pisa and INFN Sezione di Pisa,\\ 
Largo Pontecorvo 3, I-56127 Pisa, Italy
}
\author{Michele Mesiti}
\email{michele.mesiti@swansea.ac.uk}
\affiliation{Academy of advanced computing, Swansea University, \\
Singleton Park, Swansea SA2 8PP, UK
}
\author{Francesco Negro}
\email{fnegro@pi.infn.it}
\affiliation{INFN Sezione di Pisa,\\ 
Largo Pontecorvo 3, I-56127 Pisa, Italy
}
\author{Francesco Sanfilippo}
\email{sanfilippo@roma3.infn.it}
\affiliation{INFN Sezione di Roma3,\\ 
Via della Vasca Navale 84, I-00146 Roma, Italy
}
\author{Sebastiano Fabio Schifano}
\email{schifano@fe.infn.it}
\affiliation{Universit\`a degli Studi di Ferrara and INFN Sezione di Ferrara,\\ 
Via Saragat 1, I-44122 Ferrara, Italy
}
\author{Giorgio Silvi}
\email{g.silvi@fz-juelich.de}
\affiliation{J{\"u}lich Supercomputing Centre, 
Forschungszentrum J{\"u}lich,\\
Wilhelm-Johnen-Stra{\ss}e, 52428 J{\"u}lich, Germany 
}
\author{Raffaele Tripiccione}
\email{tripiccione@fe.infn.it}
\affiliation{Universit\`a degli Studi di Ferrara and INFN Sezione di Ferrara,\\
Via Saragat 1, I-44122 Ferrara, Italy
}

\date{\today}

\begin{abstract}
We investigate the fate of the Roberge-Weiss endpoint transition
and its connection with the restoration of chiral symmetry as the 
chiral limit of $N_f = 2+1$ QCD is approached.
We adopt
a stout staggered discretization on lattices with $N_t = 4$
sites in the temporal direction; the chiral limit
is approached 
maintaining a constant
physical value of the strange-to-light mass ratio and exploring 
three different light quark masses, 
corresponding to pseudo-Goldstone pion masses 
$m_\pi \simeq 100, 70$ and 50 MeV around the transition.
A finite size scaling analysis provides evidence that the 
transition remains second order, in the 3D Ising universality class,
in all the explored mass range. The residual chiral symmetry 
of the staggered action also allows us to investigate 
the relation between the Roberge-Weiss endpoint transition
and the chiral restoration transition as the 
chiral limit is approached:
our results, including the critical scaling of the 
chiral condensate, are consistent with a coincidence of the two transitions
in the chiral limit; however we are not able to discern
the symmetry controlling the critical behavior,
because the 
critical indexes relevant to the scaling of 
the chiral condensate are very close to each other 
for the two possible universality classes (3D Ising or $O(2)$).
\end{abstract}

\maketitle

\section{Introduction}
\label{intro}

Numerical investigation of QCD or QCD-like theories in the presence of 
imaginary chemical potentials coupled to quark number operators
has been the subject of various lattice studies~\cite{alford,lomb99,fp1,dl1,azcoiti,chen,Wu:2006su,NN2011,giudice,ddl07,cea2009,alexandru,cea2012,Karbstein:2006er,cea_other,Conradi:2007be,sanfo1,Takaishi:2010kc,cea_hisq1,corvo,nf2BFEPS,bellwied,gunther,gagliardi,Bornyakov:2017upg,andreoli}.
The main source of interest is the possibility of obtaining 
information about QCD at finite baryon density via analytic 
continuation, thus partially avoiding 
the sign problem. Moreover, 
numerical results at imaginary $\mu$ are also a relevant
test bed for effective models trying to reproduce
the properties of QCD at finite density~\cite{effective,takaha,Greensite:2017qfl}.
Furthermore, imaginary chemical potentials are an interesting 
extension of the QCD phase diagram per se, as, for 
particular choices of the chemical potentials, one recovers
exact symmetries even in the presence of finite quark masses,
leading to the presence of interesting phase transitions
and critical points which, in principle,
could be relevant also for the physical region of the 
phase diagram.

\begin{figure}[t!]
\includegraphics[width=0.92\columnwidth, clip]{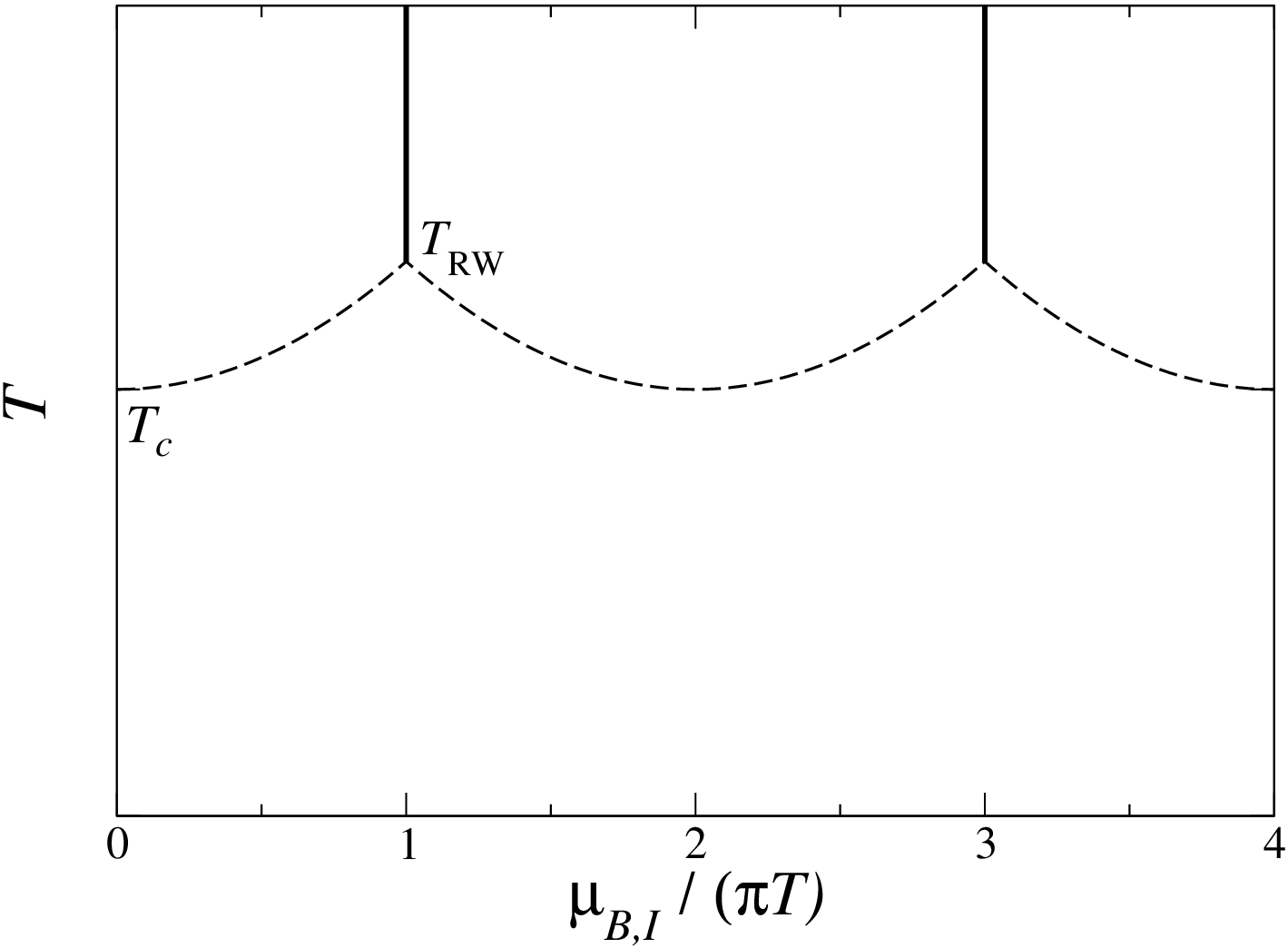}
\caption{Sketch of the phase diagram of QCD in the $T - \mu_{B,I}$
plane. The vertical lines are the RW transitions, the 
dashed lines are the analytic 
continuation of the pseudo-critical line.}\label{figrw}
\end{figure}

A well known example is QCD with an imaginary baryon chemical potential $\mu_B$,
which, for particular values,  known as Roberge-Weiss (RW) points~\cite{rw}
($\mu_B \equiv  i \mu_{B,I} = 
i k \pi T$ where $k$ is an odd integer),
has an exact $Z_2$ symmetry, which is a remnant of the original
$Z_3$ symmetry present in the pure gauge case; this simmetry gets
spontaneusouly broken at a critical temperature $T_{RW}$ 
which fixes the endpoint (RW endpoint) of first order transition
lines which are present (at fixed $\mu_B$) in the high-$T$
region of the phase diagram, as sketched in Fig.~\ref{figrw}.
 More exotic combinations
have been also considered, like those in which an exact 
$Z_{N_c}$ center
symmetry is recovered ($N_c$ being the number of colors)
by locking it to flavor symmetry
in the presence of $N_f = N_c$ degenerate 
flavors~\cite{ZN_1,ZN_2,ZN_3,ZN_4,ZN_5,ZN_6,ZN_7}.

The RW transition lines and their endpoints have been
thoroughly investigated by lattice 
simulations~\cite{fp1, dl1, ddl07, cea2009, FMRW, CGFMRW, OPRW, cea2012,PP_wilson, alexandru, wumeng, wumeng2, nf2BFEPS, nagata15, makiyama16, cuteri, nf2PP,kashiwa,
rw_physicalpoint,andreoli}
and effective models~\cite{model-rw,
Sakai:2009dv, sakai2, sakai3, sakai4, holorw, holorw2, morita, weise, pagura, 
buballa, kp13, rw-2color}. 
Early studies, performed on lattices with
$N_t = 4$ sites in the temporal
direction and using unimproved staggered fermions, 
have shown interesting features for the 
RW endpoint transition for both 
$N_f = 2$ and $N_f = 3$ degenerate 
flavors: the transition is first
order for small quark masses, likely down to the chiral limit, 
second order for intermediate masses, and 
first order again for large quark masses; 
the three regions are separated by 
tricritical points~\cite{FMRW, CGFMRW, OPRW};
for $N_f = 2$ the tricritical point delimiting
the first order chiral region takes place for
$m_\pi \simeq 400$~MeV~\cite{CGFMRW}. These results,
which suggest a strict relation 
of the RW endpoint transition 
to the chiral properties of the theory,
have been confirmed by simulations employing standard 
Wilson fermions, even if with indications 
of a strong cut-off dependence for the location
of the tricritical points: indeed, the chiral tricritical 
light pion mass has been located at 
$m_\pi \simeq 910$~MeV for $N_t = 4$ 
and at $m_\pi \simeq 670$~MeV for $N_t = 6$~\cite{cuteri}.

A systematic study adopting stout improved staggered fermions
has been reported in Ref.~\cite{rw_physicalpoint}
for $N_f = 2+1$ QCD with physical quark masses,
employing lattices with $N_t = 4,6,8$ and $10$,
i.e.~going down to lattice spacings of the order of 
$0.1$ fm. This has permitted to obtain a reliable 
continuum extrapolation for the endpoint transition
temperature, $T_{RW} \simeq 208(5)$~MeV, corresponding
to $T_{RW} / T_c \simeq 1.34(7)$ where $T_c$ is the 
pseudo-critical chiral crossover temperature at zero baryon 
chemical potential. A finite size scaling (FSS) analysis
has been performed only for $N_t = 4$ and $6$ and has 
provided evidence for a second order transition,
meaning that the chiral tricritical pion mass,
if any, is lower than the physical pion mass,
$m_\pi \simeq 135$ MeV. 
\\

In this study we extend the analysis of the RW endpoint
adopting the same improved discretization 
already used in Ref.~\cite{rw_physicalpoint}, 
exploring lower than physical quark masses,
going down to a pseudo-Goldstone pion mass of the 
order of 50~MeV.
Our purpose is twofold. First, in view of the apparent strong
reduction of the chiral first order region,
we would like to understand if a chiral tricritical 
pion mass can still be located.
The exploration of the QCD phase diagram
at zero chemical potential, 
usually summarized in the so-called Columbia plot,
has provided evidence for a general shrinking
of the first order regions as the continuum limit is approached,
and presently it is not even clear if 
a first order survives in the chiral 
limit for the $N_f = 4$ or $N_f = 3$ case~\cite{nf3_1st},
where standard universality arguments would predict 
it~\cite{piswil}. Therefore, the fact that a first 
order RW endpoint transition is still found in the chiral
and continuum limit of $N_f = 2+1$ QCD is not guaranteed.

Let us say right from the beginning that the task
itself is highly non-trivial. Indeed, due to well known
problems in the lattice discretization of fermion degrees
of freedom, a reliable approach to the chiral 
limit is only possible if the continuum limit 
is approached first~\cite{Bernard:2006vv}, while keeping 
finite size effects under control.
In other words, for a reliable investigation 
one should guarantee at the same
time that: {\em i)} 
one gets close enough to the chiral limit; {\em ii)} one stays
close enough to the continuum limit so that the chiral properties
of dynamical fermions are effective (in the present context
of staggered fermions, that means that taste symmetry breaking
is negligible and all pions becomes effectively light); 
{\em iii)} the physical 
volume of the system is still large enough, in particular
$L m_\pi \gg 1$ as $m_\pi \to 0$. Satisfying all these criteria
is still an unbearable task, even for present computational resources.

The present study is limited to $N_t = 4$ lattices and
therefore represents just a small step.
We anticipate that we have not been able to detect
any signal of a first order transition down to 
$m_\pi \simeq 50$~MeV. On one hand, one might 
consider this result as inconclusive for the reasons exposed above:
as we will show, our approach to the chiral limit
on $N_t = 4$ actually means that just one pion
mass goes to zero, while the others stay finite
and quite heavy (larger than 400 MeV), so that it is not clear 
which kind of ``chiral'' theory one is really approaching.
Nevertheless, on the other hand, it is a striking fact that
results change so drastically 
with respect to earlier
results on $N_t = 4$ lattices~\cite{FMRW, CGFMRW, OPRW}
(a tricritical pion mass
decreasing by at least one order of magnitude or vanishing at all), 
by just improving the discretization of the theory.
\\

The second purpose of our investigation is an improved understanding of
the relation between chiral and center symmetry 
and the RW transition. In the massless limit,
one could expect two different transitions temperatures, $T_\chi$ 
and $T_{RW}$, along the RW 
chemical potentials ($\mu_B = i k \pi T$ with $k$ odd),
one corresponding to the restoration of chiral symmetry
and the other to the breaking of the remnant 
center symmetry. Examples where the chiral restoration
transition is well decoupled from the center symmetry breaking transition
are well known in the literature, like for instance  
QCD with fermions in the adjoint representation~\cite{adjoint1}.
In this case, results obtained at finite quark mass
show that the two transitions are generically
close to each other; however what happens in the chiral
limit, where both symmetries are exact, is unknown.

In this case
the task is more feasible.
Indeed, even at finite lattice spacing, the staggered discretization
provides a remnant of the chiral symmetry which becomes
exact as the bare quark mass is extrapolated to zero:
it corresponds to a single generator of the original chiral group,  
it breaks spontaneously at low temperature, leading to a single massless
pion, and it gets restored at the chiral transition temperature
$T_\chi$. Therefore, it makes sense to investigate
the relation between $T_{RW}$ and $T_\chi$ in the chiral 
limit also for finite values of $N_t$, even if of course
the answer itself could be $N_t$-dependent.
Whether the two transitions coincides and,
in this case, which symmetry controls the critical behavior,
is a clear-cut question which can and should be answered.
\\

The paper is organized as follows. In Section~\ref{setup} we review the general
properties of QCD in the presence of imaginary chemical potentials,
illustrate the lattice discretization adopted for our study and give
details on our numerical setup and analysis.  
In Section~\ref{sec:fss} we report our numerical results
regarding the order of the transition for different
values of the bare quark mass, discussing also the 
corresponding values of the pion masses (pseudo-Goldstone and not) 
and the quality of our approach to the chiral limit.
In Section~\ref{chiralsymmetry} we investigate the relation
between $T_\chi$ and $T_{RW}$ as the chiral limit is approached.
Finally in Section~\ref{conclusions} we present our concluding remarks.

\begin{table}[t!]
\begin{tabular}{|c|c|c|}
\hline
$a m_l$ & $\beta$ &  $L_s^3\times 4$ lattices\\
\hline
0.003 & 3.3900  &  16 \\
      & 3.3950  &  16 \\
      & 3.4000  &  16, 20, 24 \\
      & 3.4050  &  16, 20, 24, 28 \\
      & 3.4080  &  28 \\
      & 3.4100  &  16, 20, 24 \\
      & 3.4110  &  28 \\
      & 3.4140  &  28 \\
      & 3.4150  &  16, 20, 24 \\
      & 3.4170  &  32 \\
      & 3.4175  &  28 \\
      & 3.4200  &  16, 20, 24, 28 \\
      & 3.4250  &  16, 20, 24 \\
      & 3.4300  &  16, 20, 24 \\
      & 3.4350  &  20, 24 \\
      & 3.4400  &  20, 24 \\
\hline
0.0015 & 3.3500  &  16 \\
       & 3.3550  &  16 \\
       & 3.3600  &  16, 20, 24 \\
       & 3.3650  &  16, 20, 24 \\
       & 3.3700  &  16, 20, 24 \\ 
       & 3.3750  &  16, 20, 24, 28 \\
       & 3.3800  &  16, 20, 24, 28 \\
       & 3.3820  &  32 \\
       & 3.3825  &  24, 28 \\
       & 3.3835  &  32 \\
       & 3.3850  &  16, 20, 24, 28, 32 \\
       & 3.3865  &  32 \\ 
       & 3.3875  &  24, 28 \\
       & 3.3900  &  16, 20, 24, 28 \\
       & 3.3925  &  24, 28 \\
       & 3.3950  &  16, 20, 24, 28 \\
       & 3.4000  &  16, 20, 24 \\ 
\hline
0.00075 & 3.3400  &  16 \\
        & 3.3450  &  16 \\
        & 3.3500  &  16 \\
        & 3.3550  &  16, 20, 24 \\
        & 3.3575  &  20, 24 \\
        & 3.3600  &  16, 20, 24, 28 \\
        & 3.3625  &  16, 20, 24, 28 \\
        & 3.3650  &  16, 20, 24, 28 \\
        & 3.3675  &  20, 24, 28 \\ 
        & 3.3700  &  16, 20, 24, 28 \\
        & 3.3725  &  20, 24, 28 \\ 
        & 3.3750  &  16, 20, 24, 28 \\
        & 3.3775  &  20, 24, 28 \\ 
        & 3.3800  &  16, 20, 24, 28 \\
        & 3.3850  &  20, 24 \\
\hline
\end{tabular}
\label{tabellone}
\caption{Simulation details for all finite temperature runs.}
\end{table}

\section{Numerical Setup}
\label{setup}

We consider a rooted stout  staggered discretization of $N_f=2+1$ QCD 
in the presence of imaginary quark chemical potentials $\mu_{f,I}$,
its partition function reads:
\begin{eqnarray}\label{partfunc}
Z &=& \int \!\mathcal{D}U \,e^{-\mathcal{S}_{Y\!M}} \!\!\!\!\prod_{f=u,\,d,\,s} \!\!\! 
\det{\left({M^{f}_{\textnormal{st}}[U,\mu_{f,I}]}\right)^{1/4}}
\hspace{-0.1cm}, \\
\label{tlsyact}
\mathcal{S}_{Y\!M}&=& - \frac{\beta}{3}\sum_{i, \mu \neq \nu} \left( \frac{5}{6}
W^{1\!\times \! 1}_{i;\,\mu\nu} -
\frac{1}{12} W^{1\!\times \! 2}_{i;\,\mu\nu} \right), \\
\label{fermmatrix}
(M^f_{\textnormal{st}})_{i,\,j}&=&am_f \delta_{i,\,j}+\!\!\sum_{\nu=1}^{4}\frac{\eta_{i;\,\nu}}{2}\nonumber
\left[e^{i a \mu_{f,I}\delta_{\nu,4}}U^{(2)}_{i;\,\nu}\delta_{i,j-\hat{\nu}} \right. \nonumber\\
&-&\left. e^{-i a \mu_{f,I}\delta_{\nu,4}}U^{(2)\dagger}_{i-\hat\nu;\,\nu}\delta_{i,j+\hat\nu}  \right] \, \quad ;
\end{eqnarray}
$\mathcal{S}_{Y\!M}$ is the tree level Symanzik improved gauge 
action~\cite{weisz,curci}
constructed in terms of the original link variables,
$W^{n\!\times \! m}_{i;\,\mu\nu}$ being
the trace of a $n\times m$ 
rectangular loop, while 
the staggered fermion matrix $(M^f_{\textnormal{st}})_{i,\,j}$ is
built up in terms of the two times stout-smeared~\cite{morning} links
$U^{(2)}_{i;\,\nu}$, with an
isotropic smearing parameter $\rho = 0.15$.

Adopting thermal boundary conditions (periodic/anti-periodic 
in Euclidean time for boson/fermion
fields), the temperature is given by $T = 1/(N_t a)$; 
we have fixed $N_t = 4$ in all simulations,
while the lattice spacing $a$ is a function 
of $\beta$ and of the bare quark masses.
In this study, contrary to Ref.~\cite{rw_physicalpoint}, where simulations
were done along a line of constant physics (LCP), i.e.~tuning bare masses
with $\beta$ (hence with the lattice spacing) in order to keep
the masses of physical states approximately equal to their experimental values,
we have decided to perform series of simulations around the phase
transitions for fixed values of the bare quark masses,
while keeping $m_u = m_d \equiv m_l$ and the
strange-to-light mass ratio fixed at its physical value,
$m_s/m_{l}=28.15$. There is a clear advantage stemming from this choice: 
since simulations only differ for the value
of the bare gauge coupling $\beta$, it is 
possible to make use of standard reweighting 
methods~\cite{FS1} in order to optimize 
the numerical effort; that was not possible in 
Ref.~\cite{rw_physicalpoint}, where also the 
weight of the fermion determinant changed from
one simulation to the other (because of the tuning
of the quark masses), making reweighting not feasible 
in practice.

In Ref.~\cite{rw_physicalpoint}, the critical
$\beta$ reported for $N_t = 4$ is 
$\beta_{RW}(N_t = 4) \simeq 3.45$,
which corresponds to $a m_l \simeq 0.00558$ according to 
the  LCP determined in Refs.~\cite{lattsp1, lattsp2, lattsp3}.
Based on that, we have decided to run simulations
for three different values of the quark masses,
namely $a m_l = 0.003$, $a m_l = 0.0015$ and 
$a m_l = 0.00075$: for each value we have located 
the pseudo-critical coupling $\beta_{RW}(a m_l,N_t)$ and 
performed a series of run at different values of $\beta$ 
around $\beta_{RW}$ which have then been used for reweighting.
In each case, simulations have been performed on lattices
$L_s^3 \times 4$, where different values of the spatial 
extent $L_s$ (in the 
range $16 \to 32$) have been considered to perform a FSS 
analysis. For some selected values of $\beta$
values for each mass we have performed numerical 
simulations also on $T \sim 0$ lattices, which 
have been used for renormalization and scale setting 
purposes. Table~\ref{tabellone} shows a complete list of our finite $T$
simulation parameters; statistics reach up to 
50K Rational Hybrid Monte-Carlo unit length trajectories 
for simulation points around the transition.

\begin{table}[t!]
\begin{tabular}{|c|c|c|c|c|}
\hline
$a m_l$ & $\beta$ &  $a$ [fm]  & $m_\pi$[MeV] & $m_\pi^{(1)}$[MeV] \\
\hline
       0.00075 &  3.340  &   0.29039(5)  &   48.23(6)    & 437(17)\\
       0.00075 &  3.370  &   0.28332(5)  &   49.40(7)  & 433(11)\\
       0.00075 &  3.400  &   0.27330(7)  &   51.07(6)   & 418(22)\\
\hline
       0.0015 &   3.36   &   0.28815(4)  &   68.58(3)   & 435(4)\\
       0.0015 &   3.385  &   0.28078(4)  &   70.27(3)   & 431(4)\\
       0.0015 &   3.42   &   0.26831(5)  &   73.25(3)   & 408(3)\\
\hline
       0.003  &   3.38   &   0.28616(4)  &   97.24(2)   & 444.5(4)\\
       0.003  &   3.415  &   0.27502(5)  &   100.86(3)   & 425(2)\\
       0.003  &   3.440  &   0.26539(12) &   104.00(6)   & 410.6(1.3)\\
\hline
\end{tabular}
\label{tabelltwo}
\caption{Scale setting determinations, obtained from zero temperature runs
performed on a $32^3 \times 48$ lattice; $m_\pi$ stands 
for the pseudo-Goldstone pion mass, while $m_\pi^{(1)}$ corresponds
to the first excited pion.}
\end{table}

\begin{figure}[t!]
\includegraphics[width=1\columnwidth, clip]{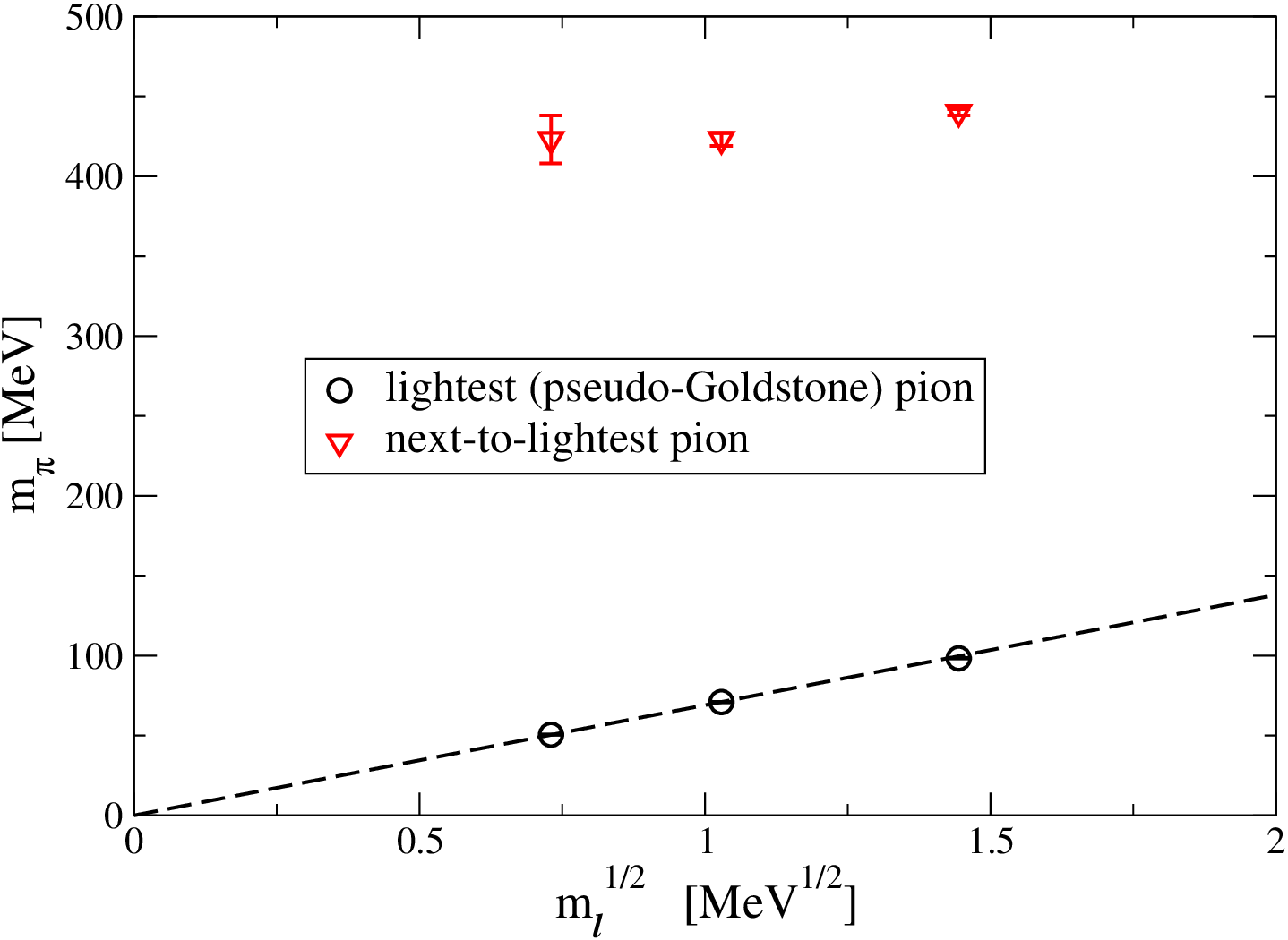}
\caption{Pion masses at $\beta=3.39$ for the three values of the bare
light mass $m_l$ we explored, expressed in physical units.
The dashed line is the result of a best fit to the 
expected $m_\pi \propto \sqrt{m_l}$ dependence.
}
\label{fig:chiral_approach}
\end{figure}

The bare quark masses have been chosen in order to 
reach, for the lowest mass, a pion mass approximately equal to 
$m_\pi = m_\pi^{(phys)} \times \sqrt{0.00075/0.00558} 
\simeq 50$~MeV around the transition point.
This estimate is only qualitative, as also 
the critical bare coupling moves as we change $a m_l$. For this 
reason, simulations at $T \simeq 0$ have been performed
in order  to obtain a direct determination of $m_\pi$ at the different
simulation points. Zero temperature runs have been used also to 
determine the lattice spacing, exploiting a technique
based on the gradient flow~\cite{wflow_1} 
and in particular the so-called $w_0$ parameter~\cite{wflow_2}.
All scale setting and 
pion mass determinations are shown 
in Table~\ref{tabelltwo}: simulations have been performed on a 
$32^3 \times 48$ lattice for all quark masses, with statistics
of the order of one thousand 
Rational Hybrid Monte-Carlo unit length trajectories 
for each simulation point.

Pion masses have been obtained
from standard Euclidean time correlators of appropriate
staggered quark operators (see, e.g., Refs.~\cite{Golterman:1984dn, Kilcup:1986dg}).
In this case, in addition to the lowest (pseudo-Goldstone)
pion state, we have also determined other pion masses,
which are expected to be higher, at finite lattice spacing,
because of the taste violations of the staggered discretization.
With the purpose of estimating the magnitude of such taste violations,
which fix the quality of our actual approach to the 
chiral limit, we report in 
Table~\ref{tabelltwo} also the value of the mass
of the first excited pion, $m_\pi^{(1)}$.

In order to better visualize the quality of our approach to 
the chiral limit, in Fig.~\ref{fig:chiral_approach} we show,
for a fixed value
of the bare gauge coupling $\beta = 3.39$, 
the values obtained for the pseudo-Goldstone pion 
and for $m_\pi^{(1)}$ as a function of the square 
root of the light bare quark mass
(in physical units).
It is quite striking that, while 
$m_\pi$ approaches zero as $m_l \to 0$ 
following quite closely the prediction of 
chiral perturbation theory, $m_\pi \propto \sqrt{m_l}$,
the first excited pion is instead 
much less affected by the change of $a m_l$.
Therefore, in our approach to the chiral limit and for what 
concerns the critical behavior around the transition,
we are effectively considering a theory with 
no more than one light pion: that is quite different 
from the physical theory and, eventually,
one would like to understand how this fact may bias the results 
obtained for the order of the phase transition.
\\

In order to implement a purely baryonic chemical potential
(i.e.~$\mu_Q = \mu_S=0$) we have set $\mu_u =
\mu_d = \mu_s \equiv \mu_q = \mu_B/3$.  An imaginary 
$\mu_q$ is equivalent to a rotation of fermionic 
temporal boundary conditions by an angle $\theta_q = {\rm Im}\,(\mu_q)/T$, 
there is therefore a periodicity in $\theta_q$, which however
is $2 \pi/N_c$ (instead of $2 \pi$) because this rotation
can be exactly canceled
by a center transformation on gauge fields.
This periodicity is smoothly realized at low $T$,
while at high $T$ the value of
$\theta_q$ selects among the three different minima of the
Polyakov loop effective potential, leading 
to first order phase transitions which occur when 
$\theta_q$ crosses 
the boundary between two adiacent center sectors.
These transitions form first order lines (RW lines)
located at $\theta_q = (2 k + 1)\pi/N_c$ 
and $k$ integer: there the average Polyakov loop 
$\langle L \rangle$ jumps from one center sector to the
other and serves as an order parameter for such transitions.
A sketch of the phase diagram is reported in Fig.~\ref{figrw}:
each RW line terminates with an endpoint located 
at a temperature $T_{\rm RW}$, where 
an exact $Z_2$ symmetry breaks spontaneously. 
Therefore, moving in temperature along these lines, one can meet
either a second order critical point in the 3D-Ising universality
class, or a first transition; in the latter case the endpoint is 
actually a triple point.

\begin{figure}[t!]
\includegraphics[width=1\columnwidth, clip]{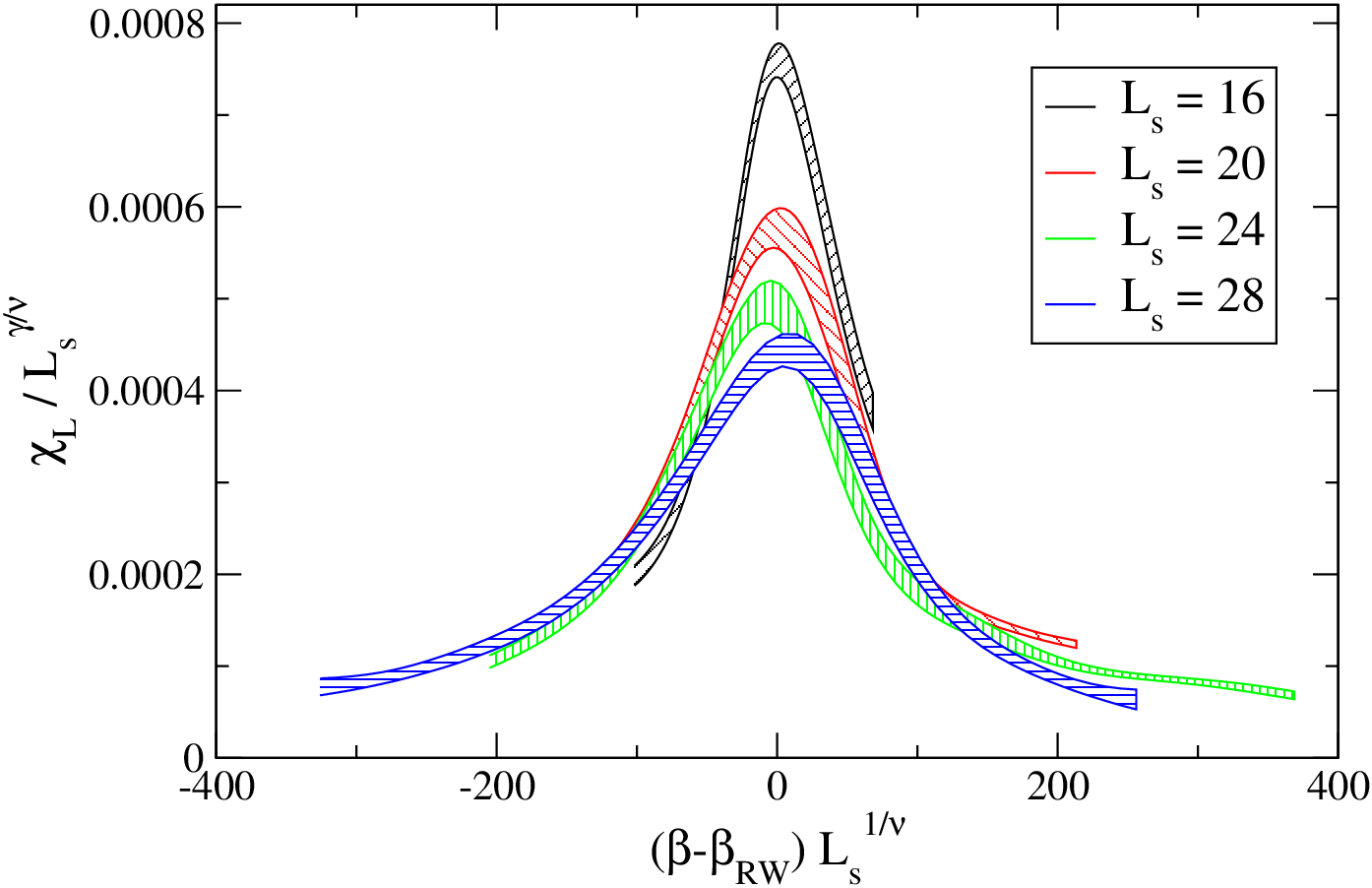}
\includegraphics[width=1\columnwidth, clip]{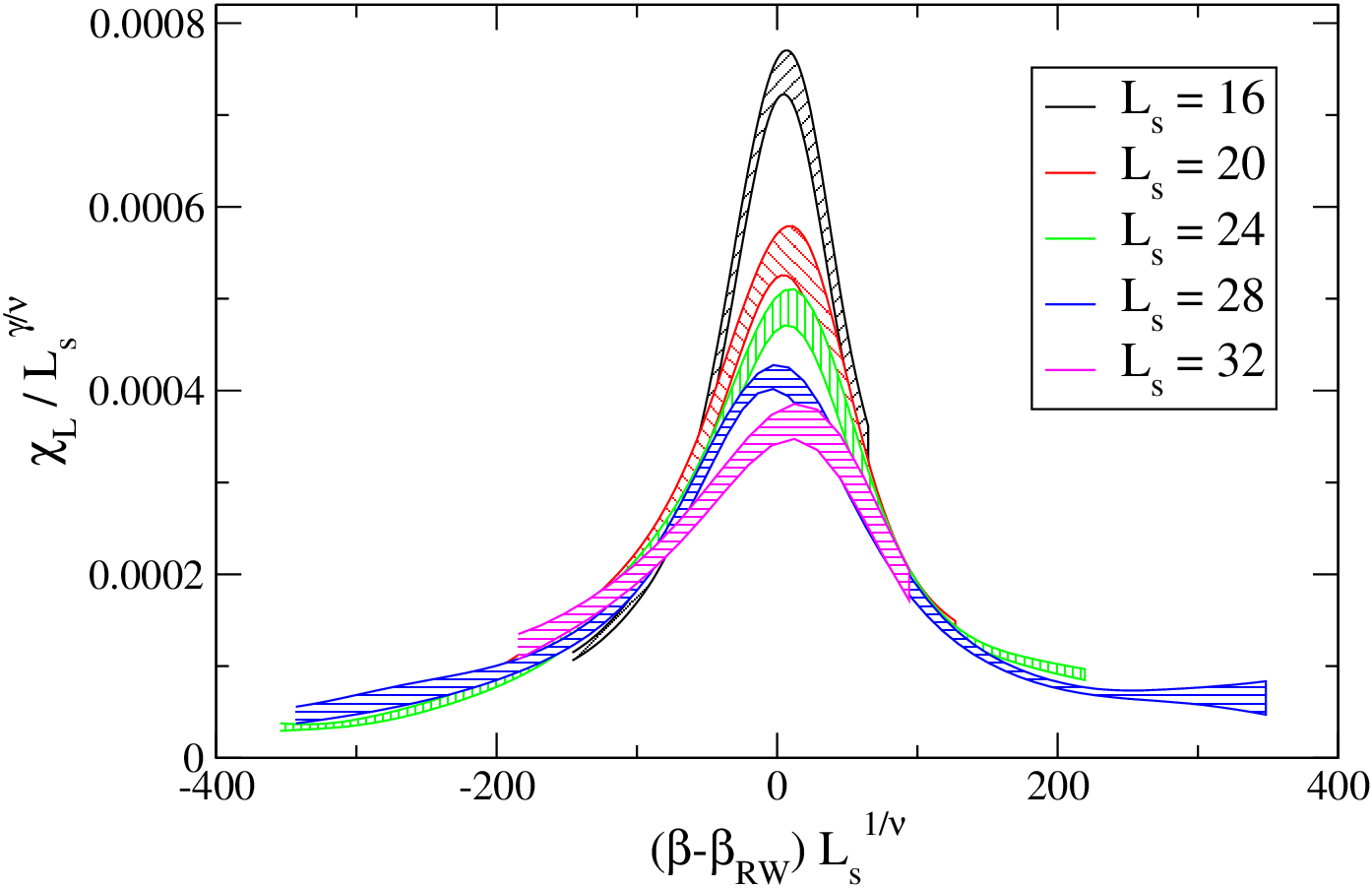}
\includegraphics[width=1\columnwidth, clip]{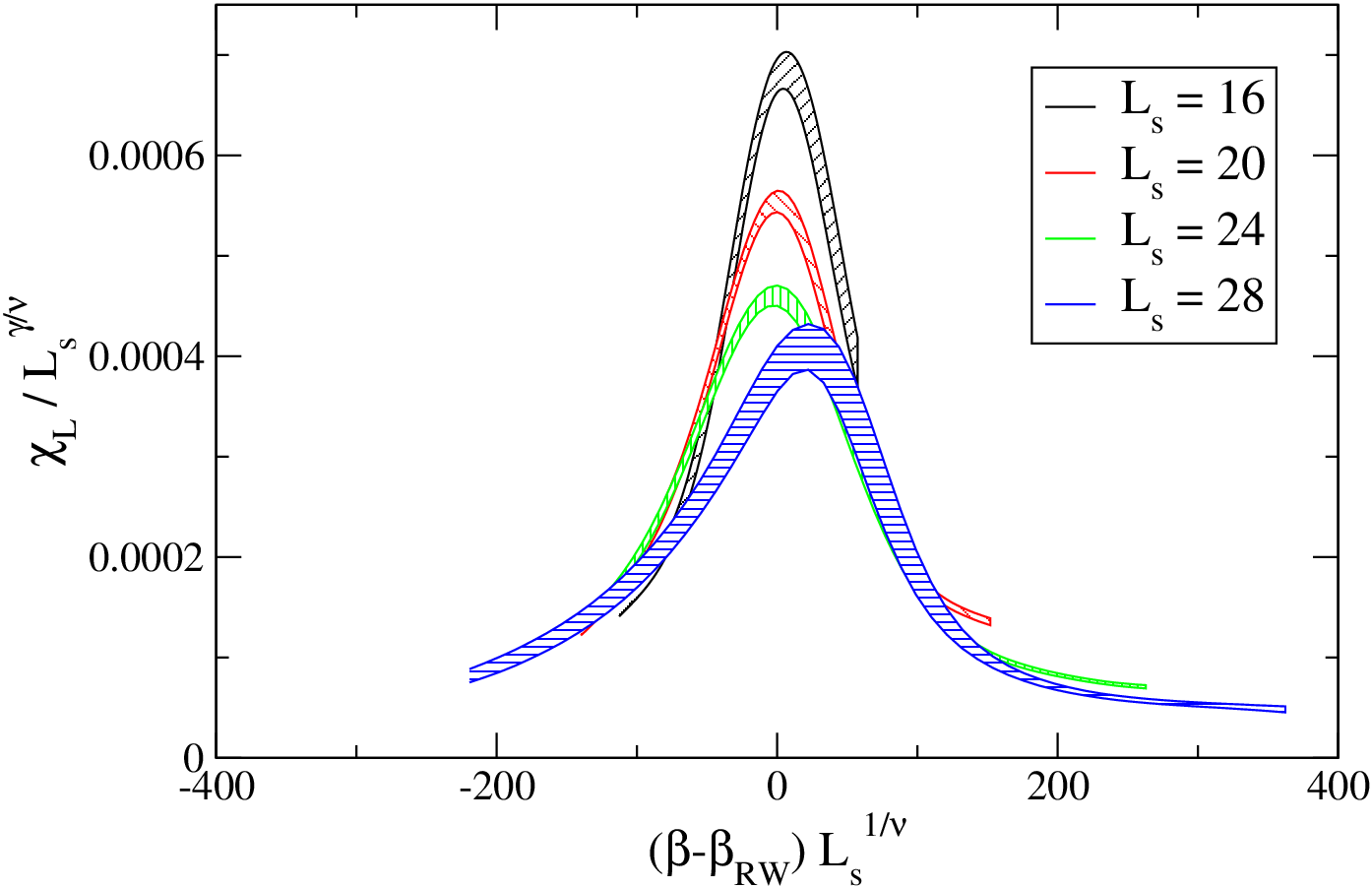}
\caption{Finite size scaling for the susceptibility of the 
Polyakov loop according to first order critical indexes. 
From top to bottom: $a m_l = 0.003$, $a m_l = 0.0015$ and $a m_l = 0.00075$. 
}
\label{fig:fss_1st}
\end{figure}

In the following we shall consider 
one particular RW line, $\theta_q = \pi$, 
for which the imaginary part of the Polyakov
loop can serve as an order parameter. In order to identify the 
universality class of the endpoint, a FSS analysis will be performed
for the susceptibility of the order parameter  
\begin{equation}\label{suscdef}
\chi_L \equiv N_t L_s^3\ (\langle ({\rm Im}(L))^2 \rangle - \langle
|{\rm Im}(L)| \rangle^2) \, ,
\end{equation}
As an alternative order parameter, one could take any of the quark number
densities (where $q=u,d,s$)
\begin{equation}
\langle n_q\rangle \equiv \frac{1}{(L_s^3 N_t)}\frac{\partial \log Z}{\partial \mu_q}
\end{equation}
which  should vanish for $\theta_q = (2 k + 1)\pi/N_c$ (because 
of the mentioned periodicity and because they are odd in $\theta_q$)
unless the $Z_2$ symmetry (which is equivalent to charge conjugation)
is spontaneously broken. However, our analysis 
will be based exclusively on the Polyakov loop.
\\

Numerical simulations 
have been performed on the COKA cluster, using
5 computing nodes, each with
$8$ NVIDIA K80 dual-GPU boards 
and two $56$~Gb/s FDR InfiniBand network interfaces.
Our parallel code (OpenStaPLE) 
is a single~\cite{gpu2} and multi~\cite{gpu3} GPU 
implementation of a standard
Rational Hybrid Monte-Carlo algorithm. It
is an evolution of a previous CUDA code~\cite{gpu1}, 
developed using the OpenACC and OpenMPI frameworks 
to manage respectively parallelism on the GPUs and among the nodes.
The multi-GPU implementation~\cite{gpu3} has been essential
in order to perform some of the zero temperature runs, which 
otherwise would have not fitted on a single GPU for memory reasons.

Of course, the most expensive simulations have been those 
regarding the lowest explored quark mass, $a m_l = 0.00075$. 
On the whole, a rough estimate of the total computational cost
of our investigation is $3 \times 10^5$ equivalent 
run-hours on a K80 GPU.

\section{Finite size scaling analysis and order of the transition}
\label{sec:fss}

The
susceptibility $\chi_L$, defined 
in Eq.~(\ref{suscdef}), is expected to scale as 
\begin{equation}\label{fss}
\chi_L = L_s^{\gamma/\nu}\ \phi (t L_s^{1/\nu}) \, , 
\end{equation}
where $t = (T - T_{\rm RW})/T_{\rm RW}$ is the reduced temperature and one has
$t \propto \beta - \beta_{RW}$ close enough to $T_{RW}$.
This means that $\chi_L/L_s^{\gamma/\nu}$, measured on different
spatial sizes, should lie on a universal scaling curve when plotted as a 
function of
$(\beta-\beta_{RW}) L_s^{1/\nu}$.  

\begin{table}[t!]
\begin{tabular}{|c|c|c|c|c|}
\hline                & $\nu$     & $\gamma$    & $\gamma/\nu$ & $1/\nu$\\
\hline $3D$ Ising     & 0.6301(4) & $1.2372(5)$ & $\sim 1.963$ & $\sim 1.587$ \\
\hline Tricritical & 1/2 & 1 & 2 & 2\\
\hline $1^{st}$ Order & 1/3       & 1           & 3            & 3\\
\hline
\end{tabular}
\caption{Critical exponents relevant to our finite size scaling
analysis (see, e.g., Refs.~\cite{LawSarb, pv_rev, isingcrit}).}\label{tab:critexp}
\end{table}

\begin{figure}[t!]
\includegraphics[width=1\columnwidth, clip]{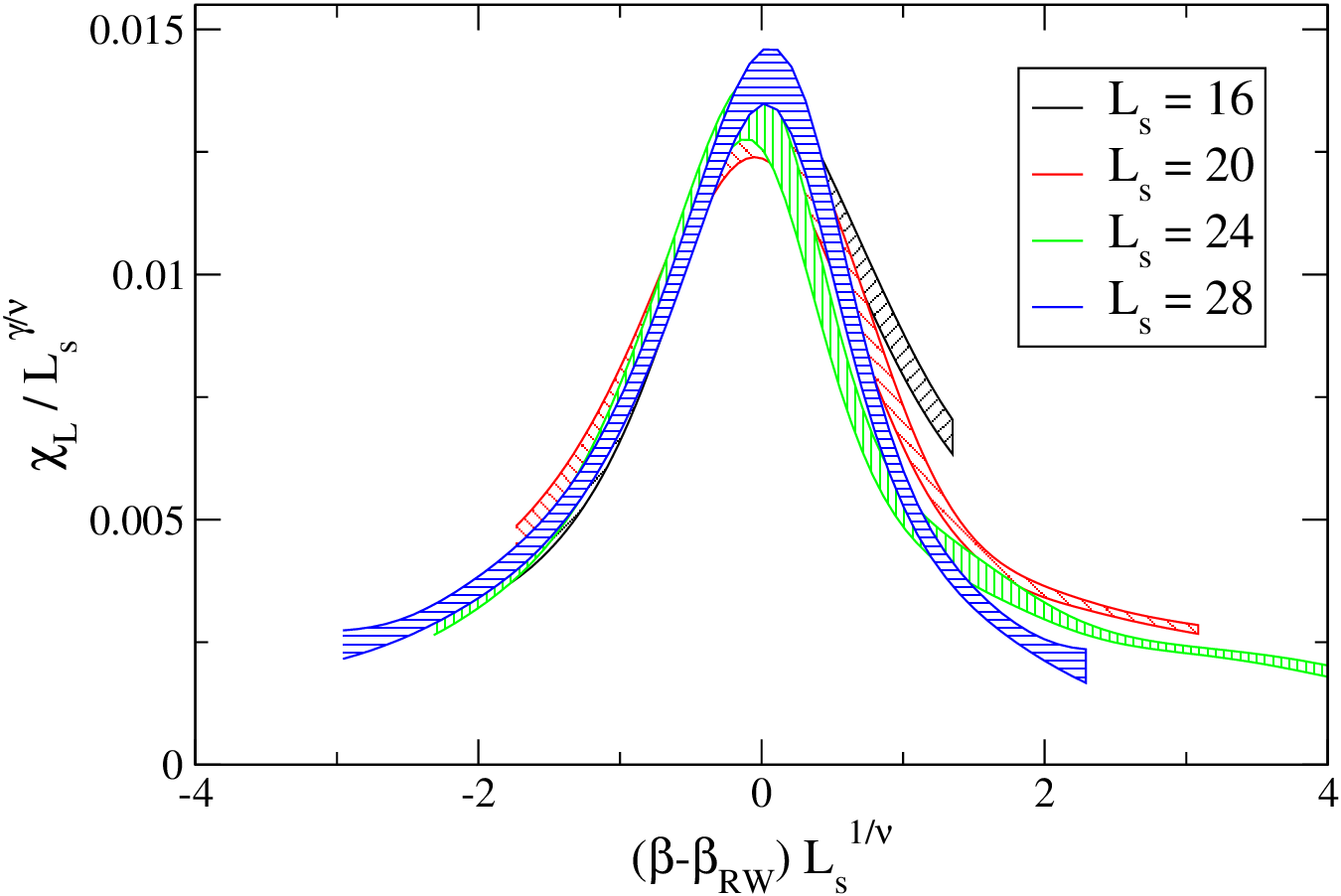}
\includegraphics[width=1\columnwidth, clip]{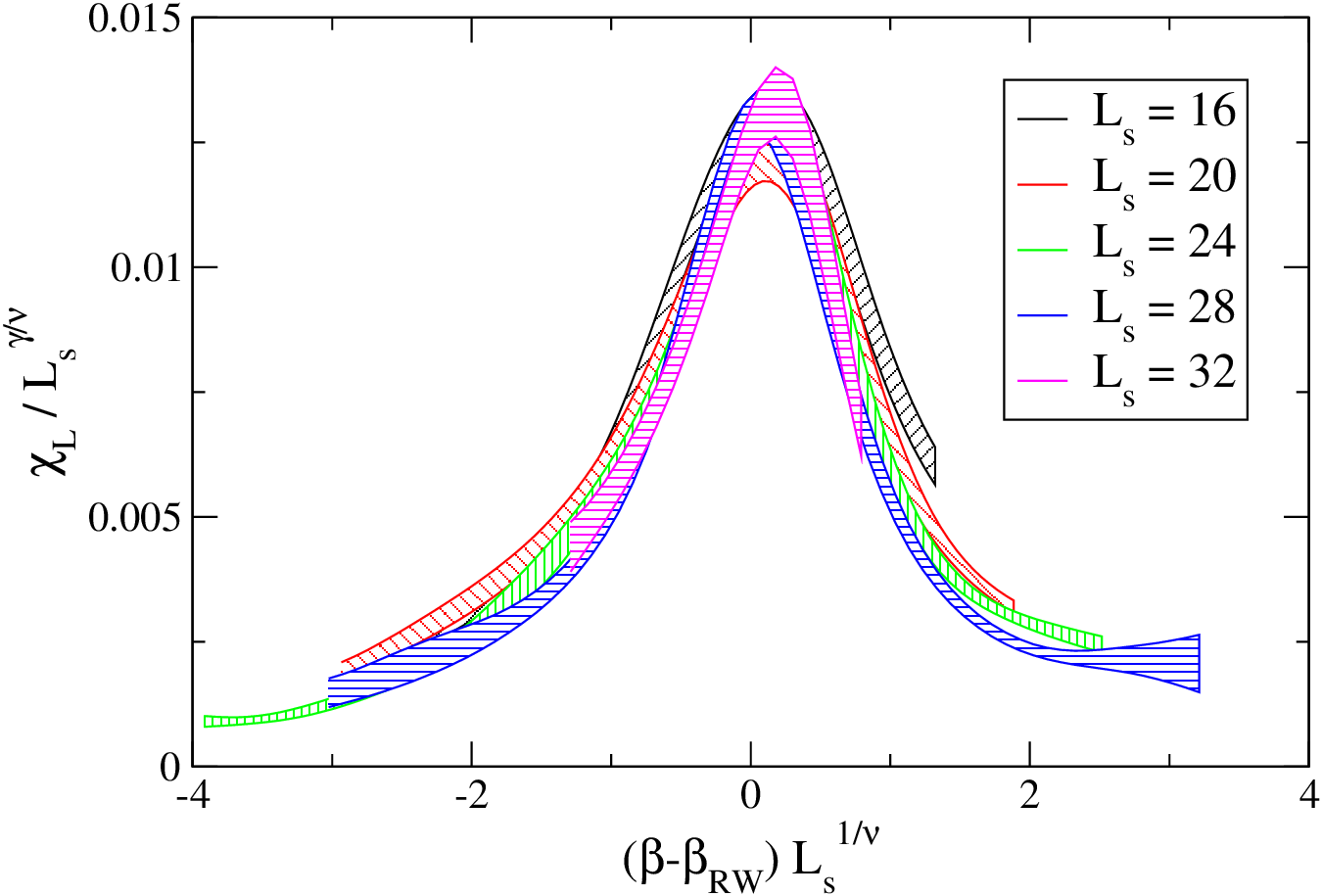}
\includegraphics[width=1\columnwidth, clip]{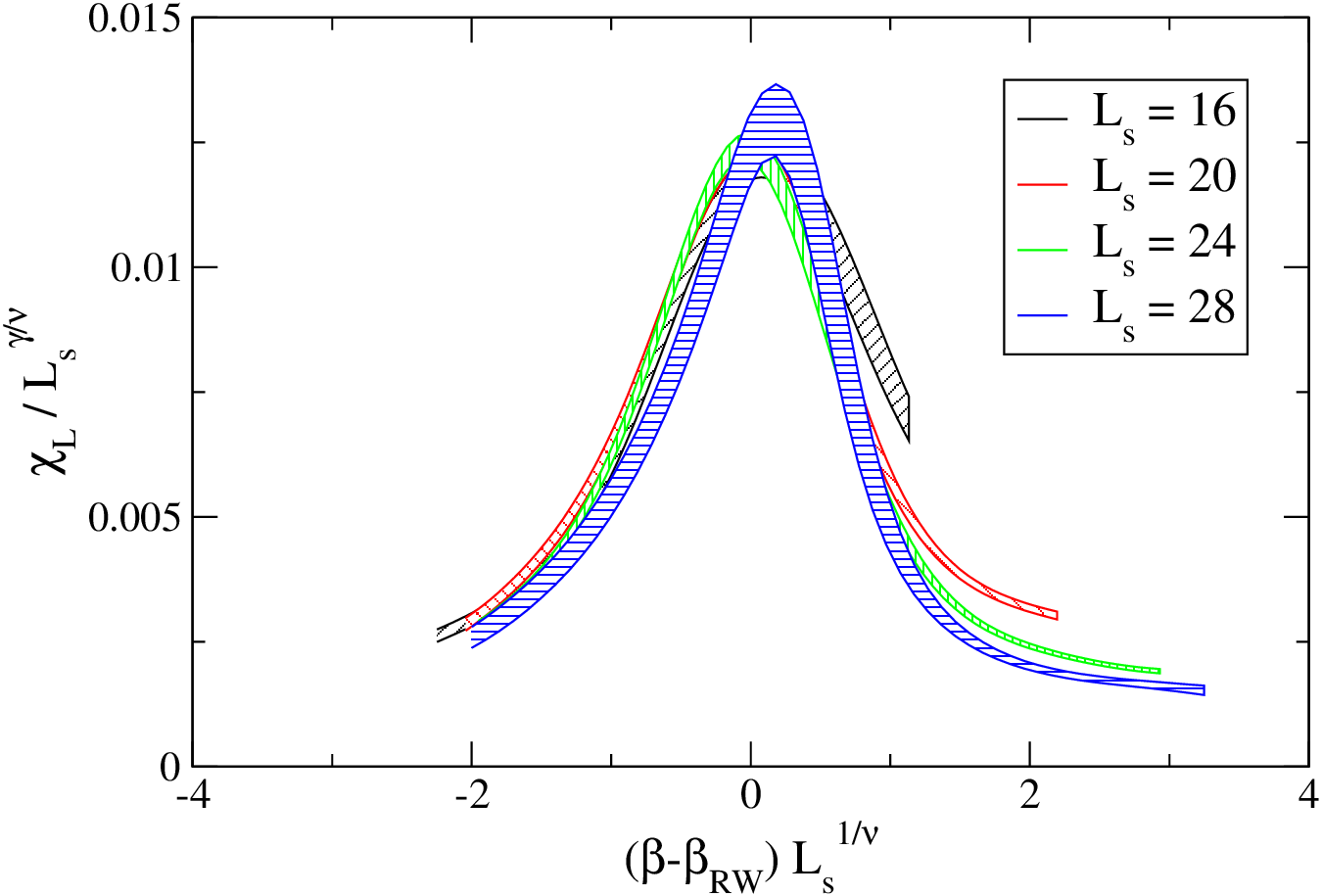}
\caption{Finite size scaling for the susceptibility of the 
Polyakov loop according to 3D-Ising critical indexes. 
From top to bottom: $a m_l = 0.003$, $a m_l = 0.0015$ and $a m_l = 0.00075$. 
}
\label{fig:fss_2nd}
\end{figure}

\begin{figure}[t!]
\includegraphics[width=1\columnwidth, clip]{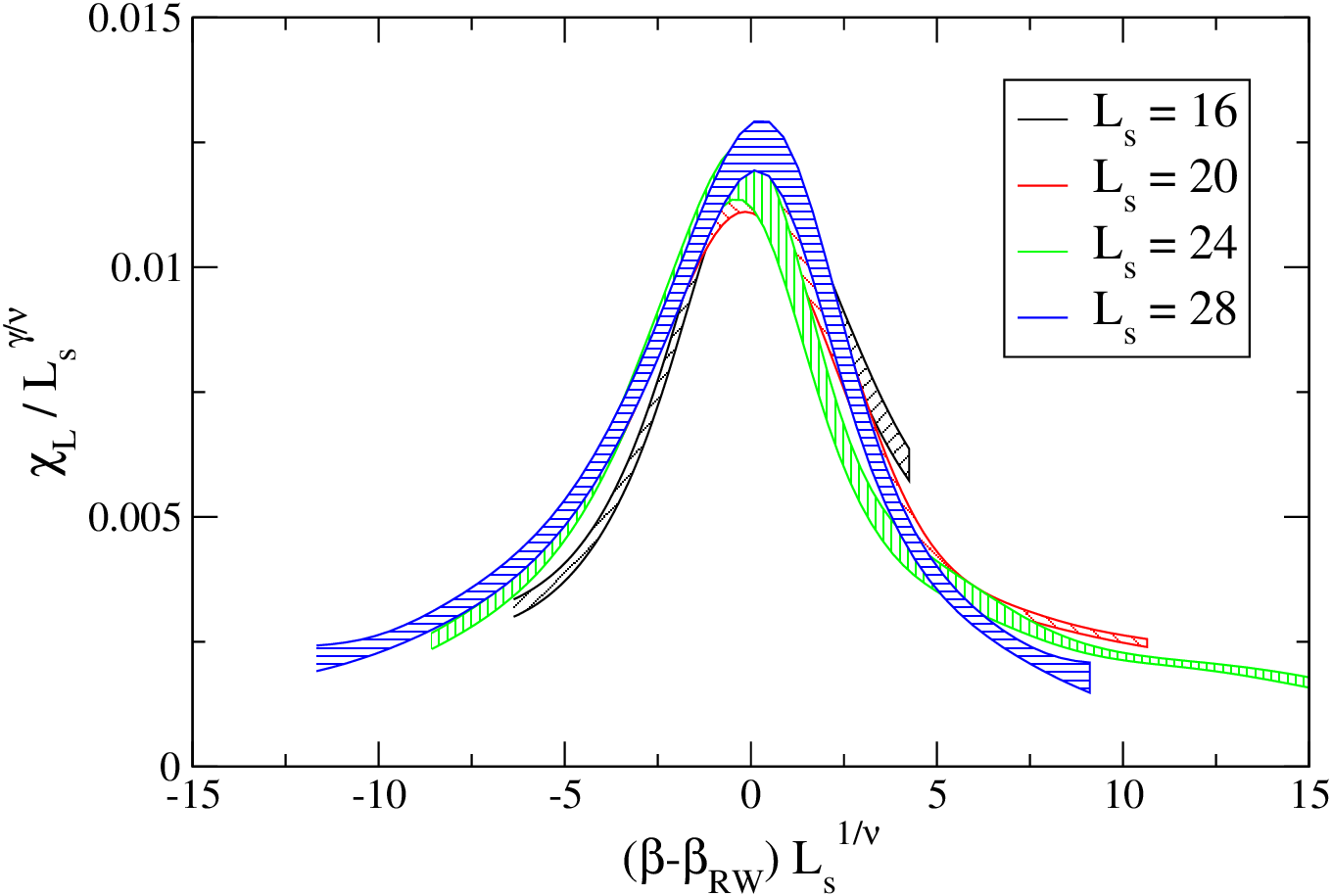}
\includegraphics[width=1\columnwidth, clip]{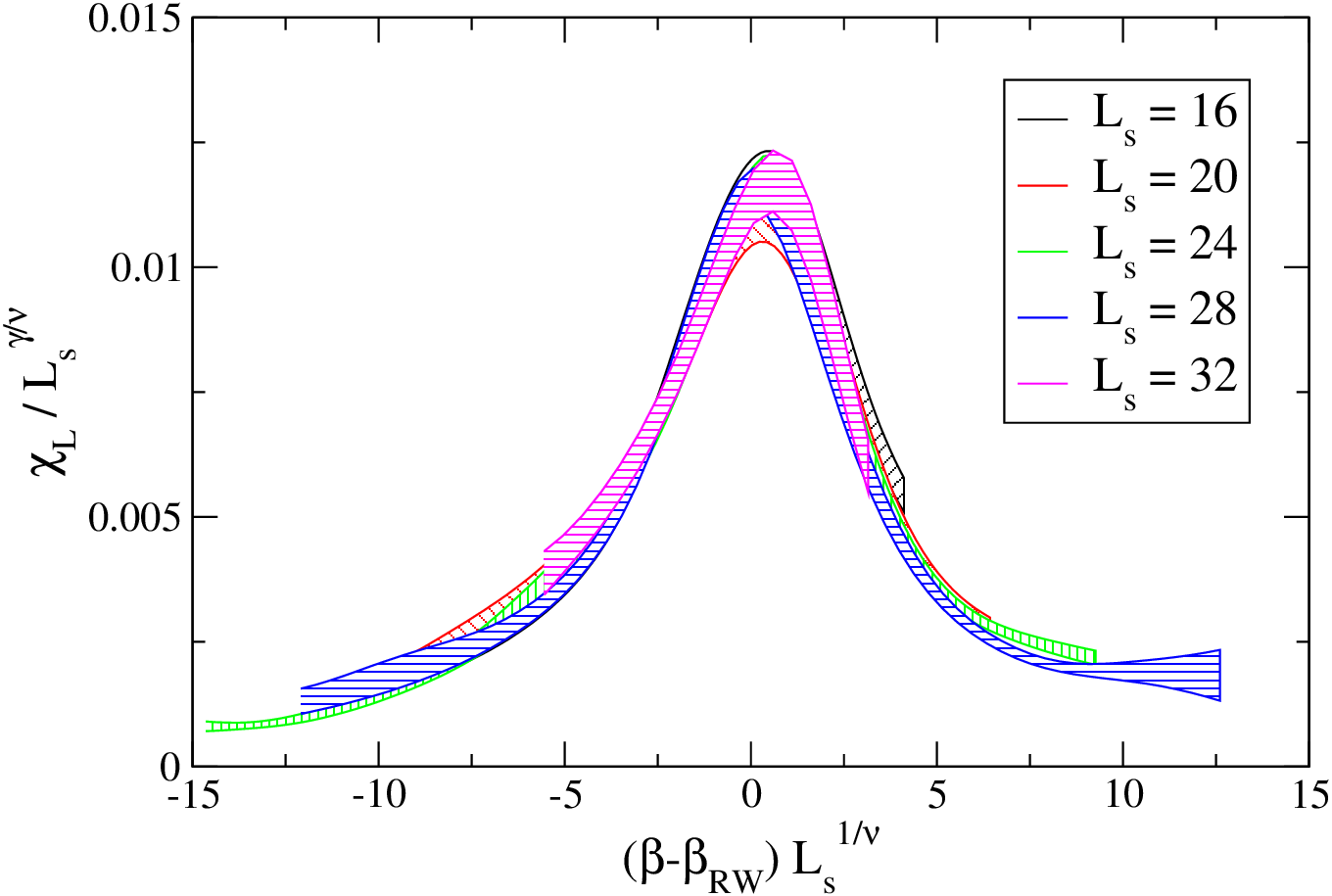}
\includegraphics[width=1\columnwidth, clip]{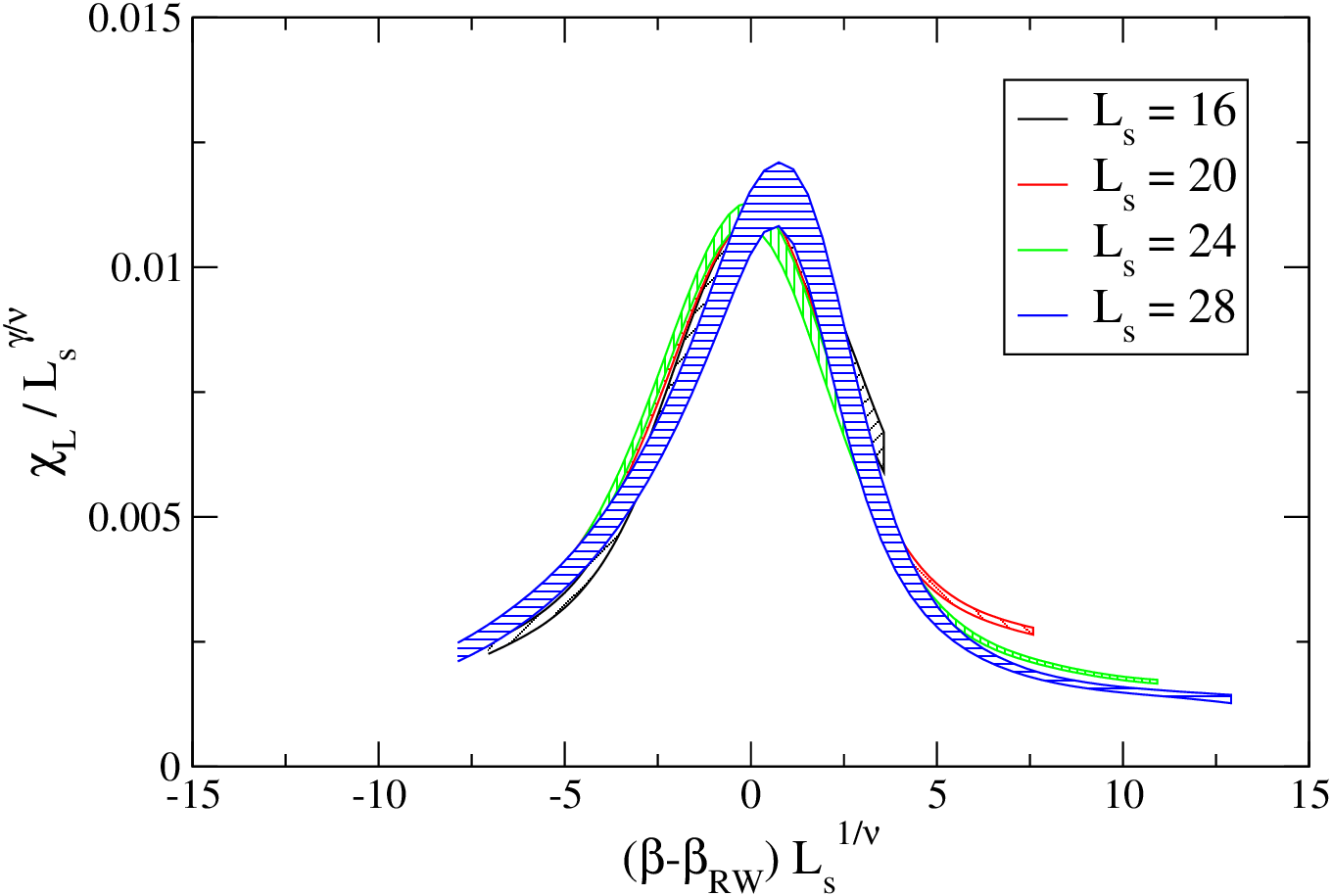}
\caption{Finite size scaling for the susceptibility of the 
Polyakov loop according to tricritical indexes. 
From top to bottom: $a m_l = 0.003$, $a m_l = 0.0015$ and $a m_l = 0.00075$. 
}
\label{fig:fss_tric}
\end{figure}

The critical exponents which are relevant to our analysis
are reported in Table~\ref{tab:critexp}. Apart from first order
and 3D-Ising exponents, we also report tricritical indexes:
they are expected to describe the critical behavior
exactly at the separation point between 
the first order and the second order region, however, before
the thermodynamic limit is really approached,
they could describe the critical behavior in a
finite neighborhood of the tricritical point~\cite{potts3d}.

A plot of $\chi_L/L_s^{\gamma/\nu}$ vs. $(\beta-\beta_{RW}) L_s^{1/\nu}$
for the three different masses is reported 
in Figs.~\ref{fig:fss_1st}, \ref{fig:fss_2nd} and 
\ref{fig:fss_tric}, respectively for 
first order, 3D-Ising and tricritical 
indexes. It clearly appears that a first
order transition is excluded for all masses,
while a reasonable scaling is obtained 
when considering both the 3D-Ising and the 
tricritical critical behavior.

\begin{figure}[t!]
\begin{center}
\includegraphics[width=0.93\columnwidth, clip]{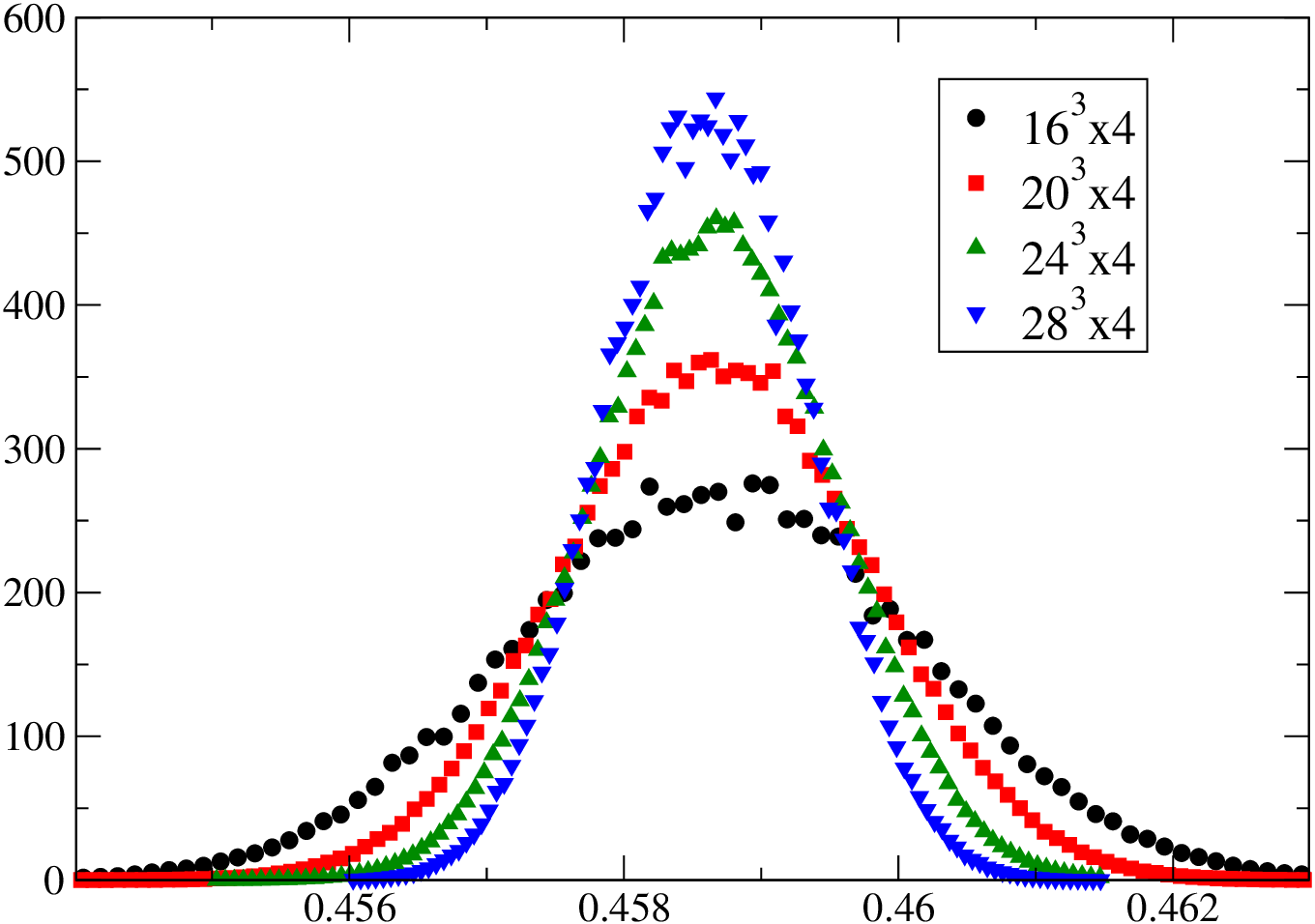}
\includegraphics[width=0.93\columnwidth, clip]{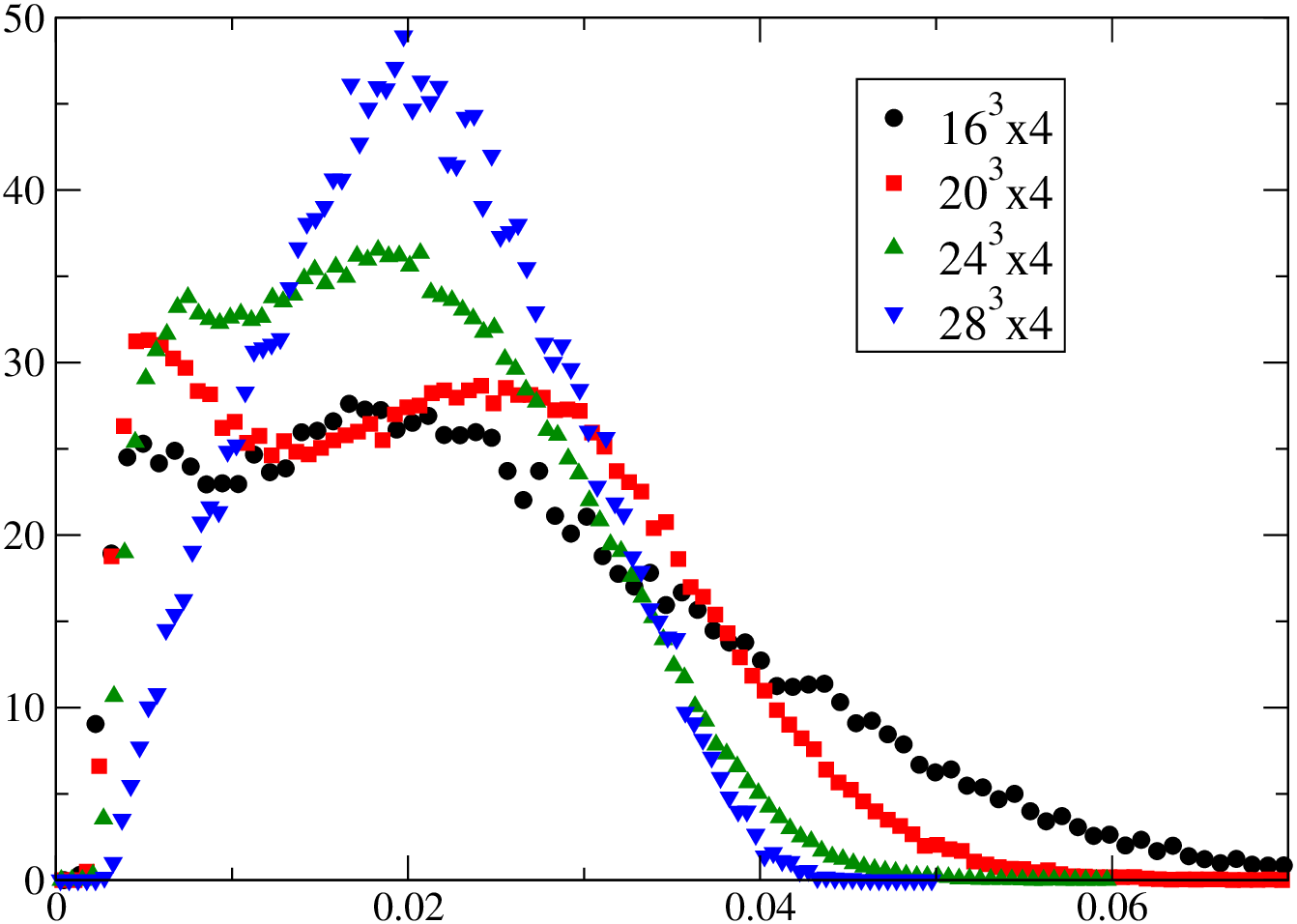}
\end{center}
\caption{Probability distribution of the plaquette (up) and of the 
unrenormalized chiral condensate (down) 
at the transition point for different
values of the spatial size $L_s$.
}
\label{histogram}
\end{figure}

As a further confirmation of the absence of a first order transition
for all explored masses, in Fig.~\ref{histogram} we report,
just for the lowest quark mass, $a m_l = 0.00075$,
the probability distribution of the plaquette and of the unrenormalized
quark condensate
at the critical point for the different lattice sizes.
A vague double peak structure is visible only
in the distribution of the chiral condensate and for small $L_s$, 
however it tends to disappear
as the thermodynamic limit is approached.

Therefore, our results suggest that a chiral first order region,
if any, is limited to a region of pion masses below 
50 MeV. There are of course many systematics that should be considered
before drawing a definite conclusions. First of all,
as we have already discussed, our approach to the chiral 
limit actually means that just one pion becomes
massless, while all other pion masses stay 
above 400 MeV. Therefore one should repeat this study 
with significantly larger values of $N_t$ (smaller lattice spacings), so that
also the other pions become lighter. In principle, 
additional chiral degrees of freedom could change 
the scenario and make the first order region 
larger, even if this is at odds with the common
experience of shrinking of first order regions
as the continuum limit is approached. Unfortunately, going to significantly 
larger values of $N_t$ is not feasible with our present 
computational resources, so this is left for future work.
 
A second remark regards the lattice sizes that we have adopted
in our study, in particular the maximum values of 
$a L_s m_\pi$ that we have reached are  
2, 3, and 4  respectively for 
$a m_l = 0.00075$, 
$a m_l = 0.0015$ and 
$a m_l = 0.003$. 
The values are not particularly large, especially
for the lowest explored quark mass.
However, we have seen no significant deviation 
from a second order scaling, and no signal for the development
of a double peak structure as the volume is increased; on the contrary,
some weak double peak signals visible in the chiral condensate distribution
for small $L_s$
have shown a tendency to disappear when going to larger
volumes.

\section{Chiral symmetry restoration and the Roberge-Weiss endpoint 
transition}
\label{chiralsymmetry}

The existence, for the staggered fermion discretization, of an
unbroken remnant of the full continuum chiral symmetry group,
permits to consider a well posed question, regarding the connection
between chiral symmetry restoration and the Roberge-Weiss transition,
even on the coarse lattices explored in our investigation.

In short, the question is the following: in the chiral limit
and for $\mu_B = i k \pi T$, with $k$ odd,
the theory enjoys both chiral symmetry and the $Z_2$ RW symmetry,
which are both expected
to undergo spontaneous symmetry breaking (or restoration)
at two temperatures, $T_\chi$ and $T_{RW}$. While results obtained
for finite quark masses indicate a generic closeness of the 
two phenomena, one would like to know if actually 
$T_\chi = T_{RW}$ or not. Moreover, if the two temperatures coincide,
which of the two symmetries dominates the transition and fixes
its universality class? The latter question is important to 
understand what are the relevant degrees of freedom around 
the transition in a non-trivial theory like QCD, where
chiral and gauge degrees of freedom are strictly entangled\footnote{
See for instance Refs.~\cite{Pelissetto:2017pxb, Pelissetto:2017sfd} 
for examples of models where the interplay with gauge degrees of freedom
can change the expected critical behavior.}.

\begin{figure}[htb!]
\includegraphics[width=1\columnwidth, clip]{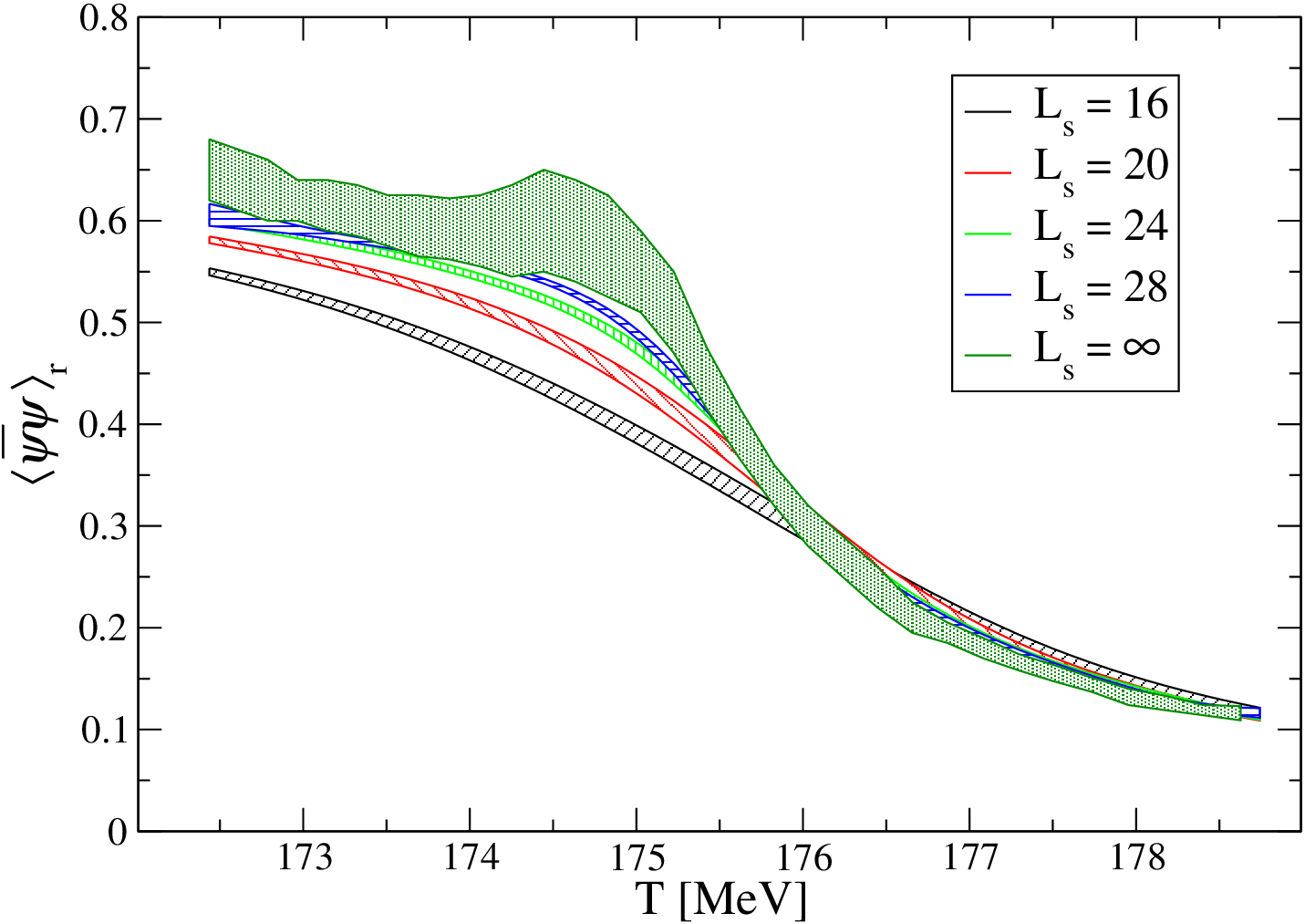}
\caption{Renormalized chiral condensate for $a m_l = 0.0015$ on different
spatial volumes and in the infinite volume limit.
}
\label{fig:termo_condensate}
\end{figure}

\begin{figure}[htb!]
\includegraphics[width=1\columnwidth, clip]{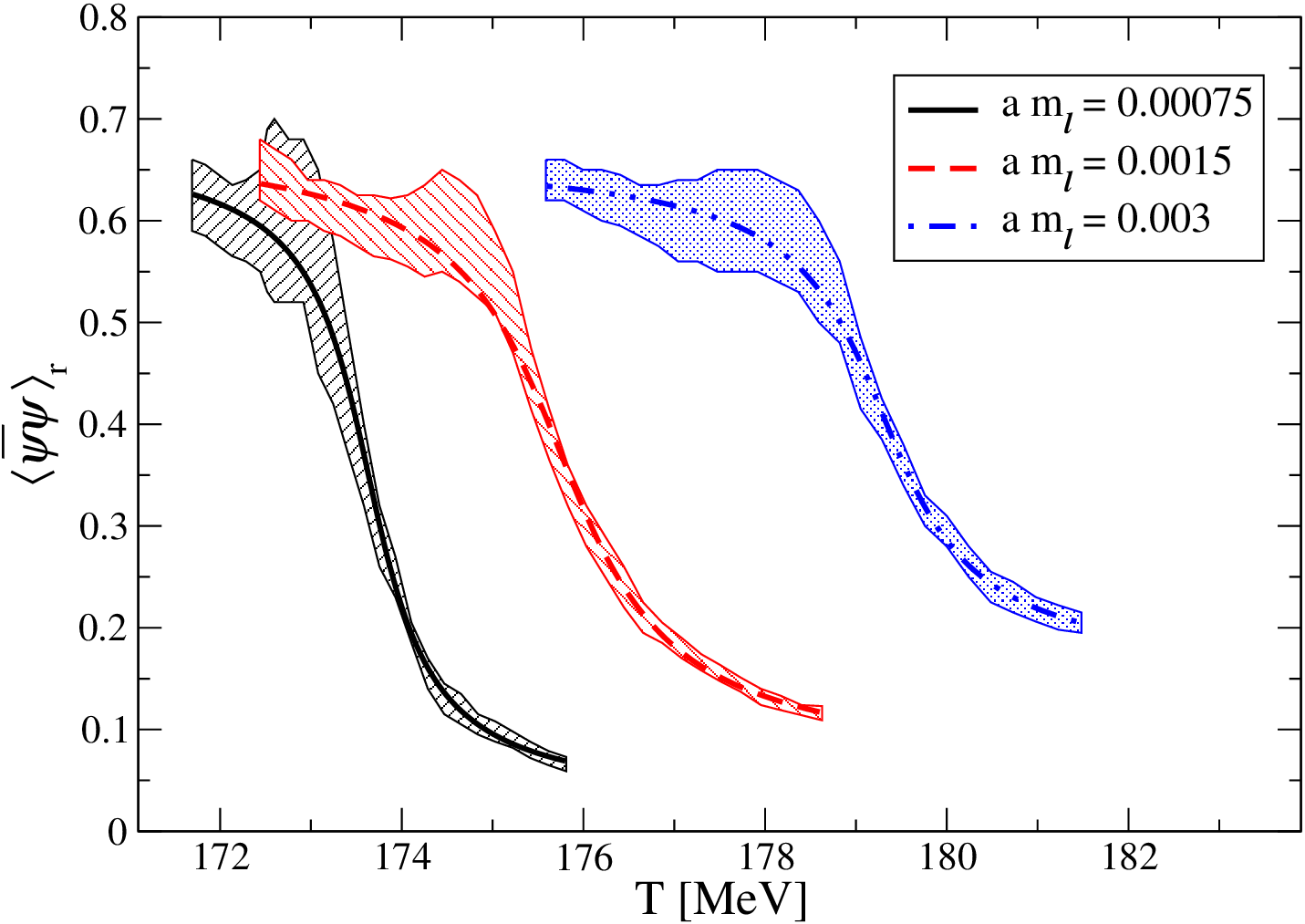}
\caption{Renormalized chiral condensate in the infinite volume limite
for the three values of the bare light quark mass explored in this study. 
The thick central lines are the result of a best fit to an $\rm {atan}$
function (see text).  
}
\label{fig:inflection_point}
\end{figure}

\begin{figure}[htb!]
\includegraphics[width=1\columnwidth, clip]{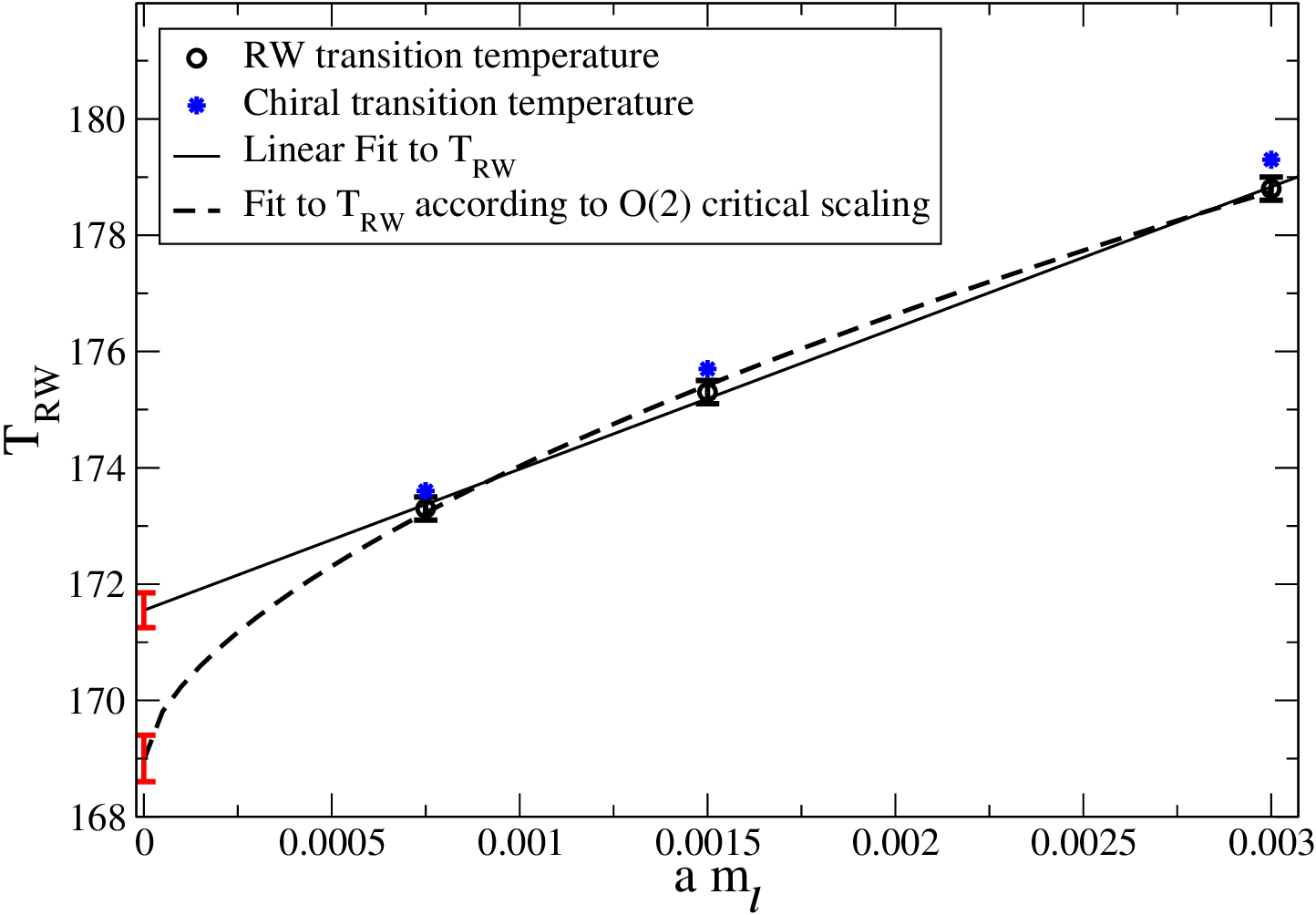}
\caption{Roberge-Weiss and chiral transition temperatures as a function
of the bare light quark mass $a m_l$. Chiral transition temperatures
are reported without error bars, which are similar to those of the 
RW temperature. Two different extrapolations to the chiral limit are provided
for $T_{RW}$, one assuming a non-critical (linear) behavior, the other
assuming an entanglement with the chiral transition and $O(2)$ critical 
indexes. The first extrapolation provides
$T_{RW}(a m_l = 0) = 171.6(4)$, the latter 
$T_{RW}(a m_l = 0) = 168.9(5)$; a similar result,
$T_{RW}(a m_l = 0) = 169.3(5)$ is obtained assuming a $Z_2$ 
critical behavior.
}
\label{fig:tc_fit}
\end{figure}

In order to answer the questions above, we consider the behavior
of the (light) chiral condensate, which is the order parameter for 
chiral symmetry breaking and is defined as follows:
\begin{equation}
\langle\bar\psi\psi\rangle_l=\frac{T}{V}\frac{\partial \log Z}{\partial m_l}=
\langle\bar{u}u\rangle+\langle\bar{d}d\rangle\ ;
\end{equation} 
where $V = L_s^3$ is the spatial volume 
and the contribution from each light flavor $f$ is expressed in terms
of the following lattice observable
\begin{equation}
    \bar{\psi}\psi_f =  \frac{1}{N_t L_s^3} \frac{1}{4} 
\tr \left[\frac{1}{M_{\rm st}^f}\right] \, ,
\end{equation}
which has been evaluated by means of noisy estimators 
(in particular up to 16 $Z_2$ random vectors have been 
used for each measurement).
The light quark condensate is 
affected by additive and
multiplicative renormalizations, which can be taken care of by, respectively,
appropriate subtractions and ratios. In particular, in this study 
we consider the following prescription~\cite{Cheng:2007jq}:
\begin{equation} \label{rencond}
\langle\bar{\psi}\psi\rangle_{r}(T)\equiv\frac{\left[
\langle \bar{\psi}\psi\rangle_l -\frac{\ds 2m_{l}}{\ds m_s}\langle \bar{s}s\rangle\right](T)}{
\left[\langle \bar{\psi}\psi\rangle_l-\frac{\ds 2m_{l}}{\ds m_s}\langle \bar{s}s\rangle\right](T=0)}\ ,
\end{equation} 
where the leading additive
renormalization, which is linear in the quark mass, 
cancels in the difference with
the strange condensate, while the multiplicative
renormalization, being independent of $T$, drops out by normalizing with respect
to quantities measured
at $T = 0$ and at the same UV cutoff. This prescription neglects 
contributions to the additive renormalization 
which are of higher order in the light quark mass; it is therefore  
particularly well suited for the present study, in which we consider
the approach to the $m_l = 0$ limit. In order to determine 
the relevant quantities at $T = 0$, we have exploited the same
set of runs already used for the determination of the 
pion masses and of the physical scale; determinations
at intermediate values of the inverse gauge coupling have been
obtained by spline interpolations.

\begin{figure}[t!]
\includegraphics[width=1\columnwidth, clip]{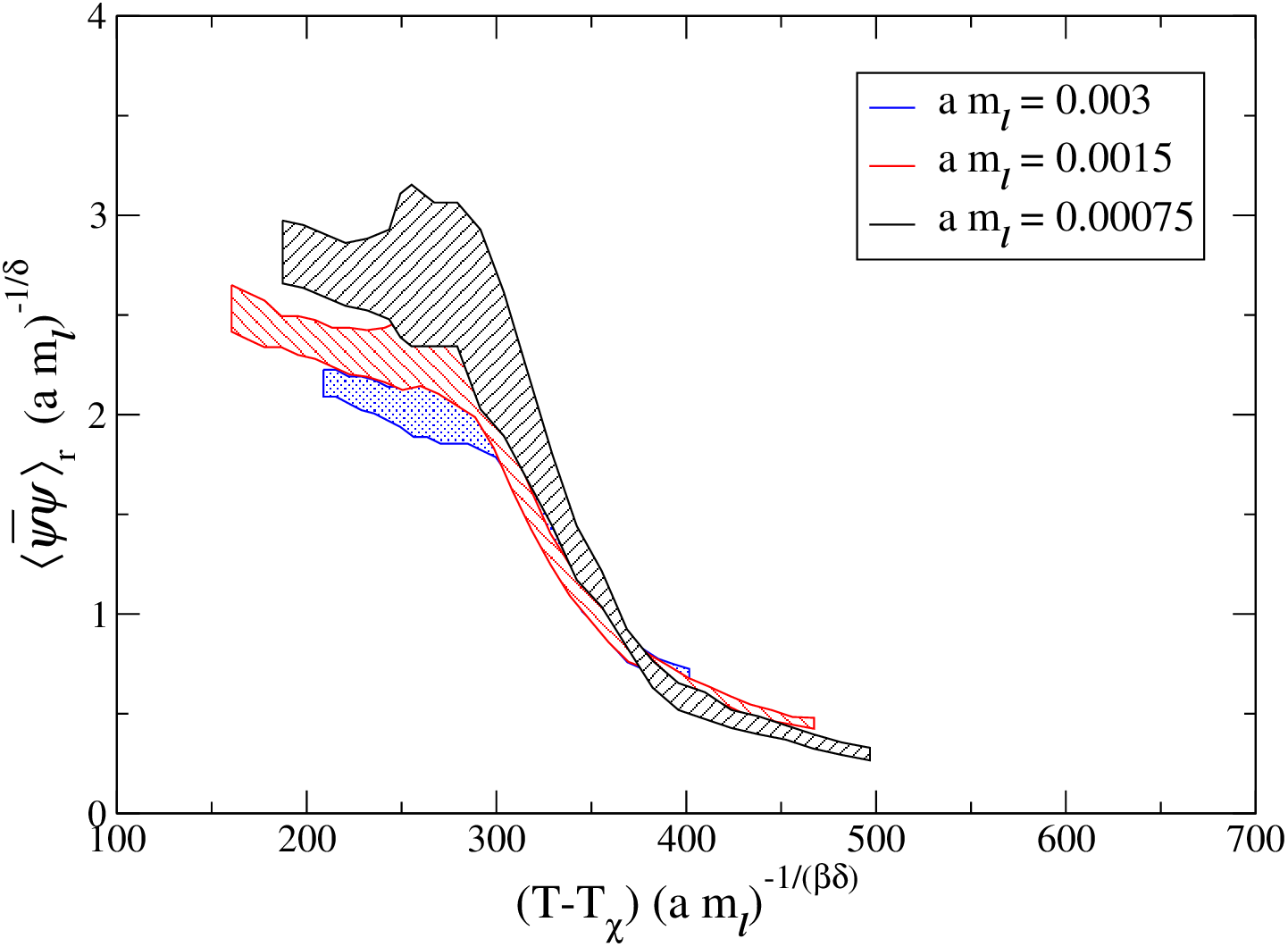}
\includegraphics[width=1\columnwidth, clip]{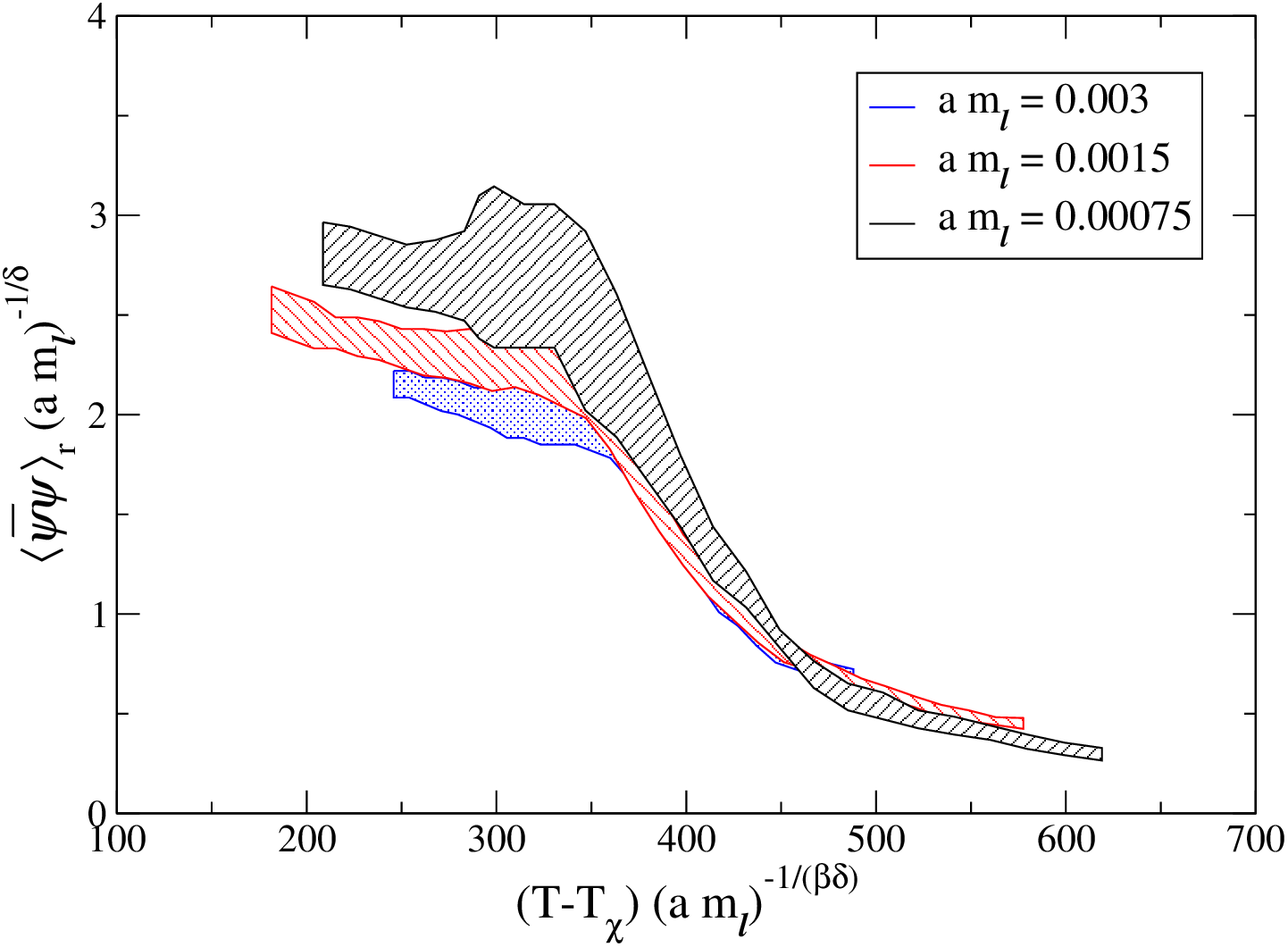}
\caption{Critical scaling around the transition temperature in the 
chiral limit for the renormalized chiral condensate. The scaling is provided
assuming two different universality classes, namely $O(2)$ 
(up) and $Z_2$
(down), in Eq.~(\ref{eos}) and fixing respectively
$T_\chi = 169.2$ and $T_\chi = 169.6$.
}
\label{fig:chiralscal}
\end{figure}

\begin{table}[t!!]
\begin{tabular}{|c|c|c|c|c|}
\hline
$a m_l$ & $T_{RW} (a m_l)$ &  $T_\chi (a m_l)$ \\
\hline
 0.00075 &  173.3(3)   &   173.6(3)  \\
  0.0015 &  175.3(2)   &   175.7(3)  \\
  0.003  &  178.8(3)   &   179.3(4)  \\
\hline
\end{tabular}
\caption{Chiral and RW critical temperatures determined for the three
different bare quark masses.}
\label{tab:crittemp}
\end{table}

An example of the renormalized chiral condensate obtained
for $a m_l = 0.0015$, and expressed as a function of 
$T$, is shown in Fig.~\ref{fig:termo_condensate}, where
determinations corresponding to different spatial extents $L_s$
are present. It is quite clear from the figure that the dependence on $L_s$
is not negligible and larger in the region around and below 
the critical temperature.
For this reason, before performing an analysis of the approach to
the chiral limit, 
we have extrapolated the chiral condensate to the infinite volume limit
at each value of the temperature.
The chiral condensate is not an order parameter for the Roberge-Weiss
transition, therefore, for finite quark mass, 
it is expected to have a smooth approach 
to the thermodynamic limit; however the degree up to which 
chiral degrees of freedom are entangled in the Roberge-Weiss transition 
is not known, moreover the behavior could be already affected
by the closeness of the chiral transition. For these reasons, the extrapolation
to the thermodynamic limit has been performed, for each temperature,
trying different fitting functions which assume either an exponential
(in $L_s$) suppression of finite size effects or power law corrections
in the spatial size: the error on the final extrapolation, which
is reported in Fig.~\ref{fig:termo_condensate} as well,
takes into account the spread among the different fitting 
ans\"atze as a source of systematic uncertainty, and is particularly more 
pronounced around and below the critical temperature.

The infinite volume extrapolations obtained for the different quark
masses are reported in Fig.~\ref{fig:inflection_point},
where data are also fitted to 
an arctangent function, 
$A = P_1 + P_2 \arctan {(P_3 (T - T_\chi(a m_l)))}$,
obtaining the values of the chiral transition temperature
$T_\chi(a m_l)$ reported in Table~\ref{tab:crittemp} and diplayed, 
together with the Roberge-Weiss critical temperatures $T_{RW}$,
in Fig.~\ref{fig:tc_fit}.
One aspect which is already clearly visibile is that 
$T_\chi(a m_l)$ is very close to $T_{RW} (a m_l)$ and,
even if it the two temperatures are actually always compatible within errors,
they seem to approach each other more closely as the chiral 
limit is approached.

This is already a good piece of evidence for the coincidence
of $T_{RW}$ and $T_\chi$ in the chiral limit. However, in order
to complete the picture, one would like to know if the drop
of the condensate at $T_\chi(a m_l)$ is actually associated
to a critical behavior around $T_\chi( a m_l = 0)$, corresponding
to the vanishing of the condensate and the restoration
of chiral symmetry at that point. Trying to answer this question,
one can also obtain information about the universality class.

The critical temperatures themselves do not provide much information.
Around the chiral transition the pseudo-critical temperatures
obtained for finite quark mass, are expected 
to scale like
\beq
T_\chi (a m_l) = T_\chi (0)  + C\cdot (a m_l)^{1/(\beta \delta)}
\label{tccrit}
\eeq 
where $\beta$ and $\delta$ are the critical indexes
of the relevant universality class. Two possibilities that we have taken
into account are the 3D $O(2)$ and $Z_2$ critical behaviors: the first 
one is naturally associated with a second order chiral transition
in the presence of just one Goldstone pion (it would be $O(4)$ in 
the continuum case, which however has practically indistiguishable critical
indexes); the second is the relevant universality class
for the RW transition and would also be associated with a critical
endpoint of a first order line present at very small quark masses.

\begin{table}[h!]
\begin{tabular}{|c|c|c|c|c|}
\hline                & $\beta$     & $\delta$    \\
\hline $3D$ Ising $Z_2$    & 0.3265(3)  & 4.789(2) \\
\hline $O(2)$ &  0.3485(2)           &  4.780(2)   \\
\hline
\end{tabular}
\caption{Critical exponents relevant to the analysis of the chiral transition (see 
Refs.~\cite{pv_rev} and \cite{critexpo2}).}\label{tab:chiralcritexp}
\end{table}

The corresponding critical indexes are reported in Table~\ref{tab:chiralcritexp}.
In Fig.~\ref{fig:tc_fit} we report best fits of $T_{RW} (a m_l)$ according 
both to the critical behavior in Eq.~(\ref{tccrit}) and to a regular
behavior $T_{RW}(a m_l) = T_{RW} (0) + C (a m_l) + O( (a m_l)^2)$.
As one can easily appreciate, even if the two ans\"atze lead to
different chiral extrapolations, they are not distinguishable 
in the quark mass range which has been actually explored and both 
fits yield acceptable values of the chi-squared test. Thus, we cannot
state, just according to $T_{RW} (a m_l)$, if the explored transition
is entangled with a chiral critical behavior as $a m_l \to 0$;
this is similar to what happens at $\mu_B = 0$, where the analysis
of $T_c (a m_l)$ alone is not enough to fix
the universality class of the chiral transition~\cite{nf2paper}.

Therefore, we turn our attention  
to the order parameter for chiral symmetry, 
i.e.~the chiral condensate, which around
a chiral transition and in the chiral restored phase
is expected to scale like~\cite{Bernard:1999fv,tchot}
\beq
\langle\bar{\psi}\psi\rangle_{r}(T, a m_l) = 
(a m_l)^{1/\delta} \phi \left( (T - T_\chi) (a m_l)^{-1/(\beta\delta)} \right) 
\label{eos}
\eeq
where $\phi$ is an appropriate scaling function.
In Fig.~\ref{fig:chiralscal} we show a plot of 
$\langle\bar{\psi}\psi\rangle_{r}(T,a m_l) (a m_l)^{-1/\delta}$
(extrapolated to the infinite volume limit)
versus  $\left( (T - T_\chi) (a m_l)^{-1/(\beta\delta)} \right)$
for both choices of critical indexes ($O(2)$ or $Z_2$).
The value of $T_\chi$ has been chosen in both cases so as 
to maximize the collapse of the condensates obtained at different values
of $a m_l$, obtaining $T_\chi = 169.2$ MeV and 
$T_\chi = 169.6$ MeV respectively for $O(2)$ or $Z_2$. Such values
are compatible with those obtained by fitting directly 
$T_\chi(a m_l)$ and the observed scaling is pretty good for both 
universality classes: this is also due to the fact that the critical
indexes $\delta$ and $\beta$ are quite similar for $O(2)$ 
and $Z_2$.

Therefore, our present results are consistent with a scenario 
in which chiral symmetry is restored exactly at $T_{RW}$
in the chiral limit, i.e.~$T_{RW} = T_\chi$. In order to distinguish 
the correct universality class one should 
explore quantities characterized by different critical
indexes, like the specific heat, which however
are less trivial to determine; in this respect,
the situation is quite similar to the present status 
of the determination of the universality class of the chiral 
transition at zero chemical potential.

Of course, our conclusions are still an extrapolation
of results obtained at finite, even if small, quark masses,
i.e.~one cannot completely exclude a priori that going to even
lighter quark masses $T_{RW}$ and $T_\chi$ separate.
Moreover, results obtained at finer lattice spacing
(i.e.~at larger values of $N_t$) could in principle be different.

\section{Discussion and Conclusions}
\label{conclusions}

We have investigated the fate of the Roberge-Weiss endpoint transition
and its relation with the restoration of chiral symmetry as the 
chiral limit of $N_f = 2+1$ QCD is approached.
The study has been performed on lattices with $N_t = 4$ 
sites in the temporal direction, a stout staggered discretization
for the fermion sector and the tree level Symanzik improved action
for the pure gauge sector. We have worked at fixed values
of the bare quark masses around the transition points, in order
to easily exploit multi-histogram methods, maintaining a 
physical strange-to-light mass ratio ($m_s/m_l = 28.15$) and exploring three 
different light quark masses, $a m_l = 0.003, 0.0015$ and
0.00075, corresponding respectively to pseudo-Goldstone pion masses 
$m_\pi \simeq 100, 70$ and 50 MeV around the transition.

The imaginary quark chemical potential has been fixed to 
$\mu_{f,I}/T = \pi$ for all flavors, so that the imaginary 
part of the Polyakov loop has been taken as an order parameter for the 
RW transition. An analysis of the finite size scaling of its
susceptibility has excluded the presence of a first order transition
for all values of the quark mass; this fact has been confirmed
by an inspection of the probability 
distribution of the plaquette and of the chiral condensate 
at the transition points, which have revelead no double peak 
structures as the thermodynamic limit is approached.
On the contrary, a good scaling has been observed according
to the predicted second order critical behavior, i.e.~that of the
three dimensional $Z_2$ (Ising) universality class.

Therefore, our results still provide no evidence of a first
order region around the chiral point 
for the RW transition, which for $N_f = 2$ 
unimproved staggered fermions was located 
below $m_\pi \simeq 400$~MeV~\cite{CGFMRW}
for $N_t = 4$. A strong cutoff dependence 
of the tricritical pion mass has been found
also in studies with Wilson fermions~\cite{PP_wilson,cuteri},
however it is striking that the tricritical
pion mass can go down at least one order of magnitude
(or disappear at all) by just improving the 
discretization at fixed $N_t$. 

Our results clearly need further refinement in some respects. 
Indeed,
because of the taste symmetry breaking of staggered fermions,
the chiral limit is approached only by the pseudo-Goldstone
pion directly linked to the residual staggered chiral symmetry,
while all the others stay above 400 MeV and do not seem
to be much affected by the $a m_l \to 0$ limit 
(see Fig.~\ref{fig:chiral_approach}). Therefore, even 
if not looking quite natural, it cannot 
be excluded apriori that, as the continuum limit is approached
and the full set of chiral degrees of freedom come into play,
the critical behavior changes and the (possible) first order
region around the chiral point enlarges again.
Unfortunately, exploring larger values of $N_t$ while
approaching the chiral limit would require computational
resources which are presently not available to us.

In spite of these caveats, thanks to the exact residual chiral symmetry of 
staggered fermions, we have been able to answer a different but related
question regarding the relation between the RW transition
and the chiral restoration transition. In the chiral limit
both symmetries are exact and predict the existence of 
a phase transition with well defined critical
temperatures, $T_{RW}$ and $T_\chi$: whether the two transitions coincides and,
in this case, which symmetry controls the critical behavior,
is a clear-cut question. Our results have shown 
that, for all explored masses, the renormalized 
chiral condensate drops sharply with an inflection point 
in coincidence (within error bars) with the 
location of RW endpoint transition; morover, the behavior
of the condensate around the transition for the different masses
and temperatures 
scales consistently with a critical behavior corresponding to 
chiral symmetry restoration in the chiral limit 
(see Eq.~(\ref{eos}) and Fig.~\ref{fig:chiralscal}). Therefore,
our results are consistent with $T_{RW} = T_\chi$.
Regarding the critical behavior, we have not been able to distinguish 
between an $O(2)$ (chiral) or $Z_2$ (Roberge-Weiss) universality class
in the chiral limit, mostly because the  critical indexes
associated with the chiral order parameter ($\delta$ and $\beta$) are 
almost coincident (in comparison with our numerical precision) in the two cases.
Of course, our present results do not exclude that the situation
might be different as the continuum limit is approached.

\acknowledgments
Numerical simulations have been performed on the COKA cluster at INFN-Ferrara. 
FN acknowledges financial support from the INFN HPC\_HTC project.


\begin{thebibliography}{99}



\bibitem{alford}
M.~G. Alford, A.~Kapustin, and F.~Wilczek, 
 Phys. Rev. D {\bf 59}, 054502 (1999)
  [hep-lat/9807039].

\bibitem{lomb99}
 M.-P. Lombardo, 
 Nucl. Phys. Proc. Suppl. {\bf 83}, 375 (2000)
 [hep-lat/9908006].

\bibitem{fp1} 
  P.~de Forcrand and O.~Philipsen,
  Nucl.\ Phys.\ B {\bf 642}, 290 (2002)
  [hep-lat/0205016];   Nucl.\ Phys.\ B {\bf 673}, 170 (2003)
  [hep-lat/0307020]; JHEP {\bf 0701}, 077 (2007)
  [hep-lat/0607017]; JHEP {\bf 0811}, 012 (2008)
  [arXiv:0808.1096 [hep-lat]].

\bibitem{dl1} 
  M.~D'Elia and M.~P.~Lombardo,
  Phys.\ Rev.\ D {\bf 67}, 014505 (2003)
  [hep-lat/0209146]; Phys.\ Rev.\ D {\bf 70}, 074509 (2004)
  [hep-lat/0406012].

\bibitem{azcoiti}
V.~Azcoiti, G.~Di Carlo, A.~Galante and V.~Laliena,
  Nucl.\ Phys.\ B {\bf 723}, 77 (2005)
  [hep-lat/0503010].

\bibitem{chen}
H.~S.~Chen and X.~Q.~Luo,
  Phys.\ Rev.\ D {\bf 72}, 034504 (2005)
  [hep-lat/0411023].


\bibitem{Karbstein:2006er} 
  F.~Karbstein and M.~Thies,
  Phys.\ Rev.\ D {\bf 75}, 025003 (2007)
  [hep-th/0610243].

\bibitem{cea_other} P.~Cea, L.~Cosmai, M.~D'Elia, A.~Papa,
  JHEP {\bf 0702}, 066 (2007)
  [hep-lat/0612018];
  Phys.\ Rev.\ D {\bf 77}, 051501 (2008)
  [arXiv:0712.3755 [hep-lat]];
  Phys.\ Rev.\ D {\bf 81}, 094502 (2010)
  [arXiv:1004.0184 [hep-lat]].



\bibitem{Wu:2006su}
L.~K.~Wu, X.~Q.~Luo and H.~S.~Chen,
  Phys.\ Rev.\ D {\bf 76}, 034505 (2007)
  [hep-lat/0611035].


\bibitem{NN2011}
K.~Nagata and A.~Nakamura,
  Phys.\ Rev.\ D {\bf 83}, 114507 (2011)
  [arXiv:1104.2142 [hep-lat]].

\bibitem{giudice}
P.~Giudice and A.~Papa,
  Phys.\ Rev.\ D {\bf 69}, 094509 (2004)
  [hep-lat/0401024].



\bibitem{ddl07} 
  M.~D'Elia, F.~Di Renzo and M.~P.~Lombardo,
  Phys.\ Rev.\ D {\bf 76}, 114509 (2007)
  [arXiv:0705.3814 [hep-lat]].

\bibitem{cea2009} 
  P.~Cea, L.~Cosmai, M.~D'Elia, C.~Manneschi and A.~Papa,
  Phys.\ Rev.\ D {\bf 80}, 034501 (2009)
  [arXiv:0905.1292 [hep-lat]].

\bibitem{alexandru}
A.~Alexandru and A.~Li,
  PoS LATTICE {\bf 2013}, 208 (2013)
  [arXiv:1312.1201 [hep-lat]].


\bibitem{cea2012} P.~Cea, L.~Cosmai, M.~D'Elia, A.~Papa and F.~Sanfilippo,
  Phys.\ Rev.\ D {\bf 85}, 094512 (2012)  [arXiv:1202.5700 [hep-lat]].

\bibitem{Conradi:2007be} 
  S.~Conradi and M.~D'Elia,
  Phys.\ Rev.\ D {\bf 76}, 074501 (2007)
  [arXiv:0707.1987 [hep-lat]].

\bibitem{sanfo1}
M.~D'Elia and F.~Sanfilippo,
  Phys.\ Rev.\ D {\bf 80}, 014502 (2009)
  [arXiv:0904.1400 [hep-lat]].

\bibitem{Takaishi:2010kc} 
  T.~Takaishi, P.~de Forcrand and A.~Nakamura,
  PoS LAT {\bf 2009}, 198 (2009)
  [arXiv:1002.0890 [hep-lat]].

\bibitem{cea_hisq1}
P.~Cea, L.~Cosmai and A.~Papa,
  Phys.\ Rev.\ D {\bf 89}, 074512 (2014)
  [arXiv:1403.0821 [hep-lat]];
  Phys.\ Rev.\ D {\bf 93}, 014507 (2016)
  [arXiv:1508.07599 [hep-lat]].

\bibitem{nf2BFEPS} 
  C.~Bonati, P.~de Forcrand, M.~D'Elia, O.~Philipsen and F.~Sanfilippo,
  Phys.\ Rev.\ D {\bf 90}, 074030 (2014)
  [arXiv:1408.5086 [hep-lat]].

\bibitem{corvo}
C.~Bonati, M.~D'Elia, M.~Mariti, M.~Mesiti, F.~Negro and F.~Sanfilippo,
  Phys.\ Rev.\ D {\bf 90}, 114025 (2014)
  [arXiv:1410.5758 [hep-lat]];
  Phys.\ Rev.\ D {\bf 92}, 054503 (2015)
  [arXiv:1507.03571 [hep-lat]].

\bibitem{bellwied}
R.~Bellwied, S.~Borsanyi, Z.~Fodor, J.~Gunther, S.~D.~Katz, C.~Ratti and K.~K.~Szabo,
  Phys.\ Lett.\ B {\bf 751}, 559 (2015)
  [arXiv:1507.07510 [hep-lat]].




\bibitem{gunther}
J.~Gunther, R.~Bellwied, S.~Borsanyi, Z.~Fodor, S.~D.~Katz, A.~Pasztor and C.~Ratti,
  EPJ Web Conf.\  {\bf 137}, 07008 (2017)
  [arXiv:1607.02493 [hep-lat]].

\bibitem{gagliardi}
M.~D'Elia, G.~Gagliardi and F.~Sanfilippo,
  Phys.\ Rev.\ D {\bf 95}, no. 9, 094503 (2017)
  [arXiv:1611.08285 [hep-lat]].

\bibitem{Bornyakov:2017upg} 
  V.~G.~Bornyakov {\it et al.},
  arXiv:1712.02830 [hep-lat].


\bibitem{andreoli} 
  M.~Andreoli, C.~Bonati, M.~D'Elia, M.~Mesiti, F.~Negro, A.~Rucci and F.~Sanfilippo,
  Phys.\ Rev.\ D {\bf 97}, no. 5, 054515 (2018)
  [arXiv:1712.09996 [hep-lat]].

\bibitem{effective}
J.~Greensite and K.~Langfeld,
  Phys.\ Rev.\ D {\bf 90}, no. 1, 014507 (2014);
Phys.\ Rev.\ D {\bf 90}, no. 11, 114507 (2014).

\bibitem{takaha} 
  J.~Takahashi, H.~Kouno and M.~Yahiro,
  Phys.\ Rev.\ D {\bf 91}, no. 1, 014501 (2015);
J.~Takahashi, J.~Sugano, M.~Ishii, H.~Kouno and M.~Yahiro
  arXiv:1410.8279 [hep-lat].

\bibitem{Greensite:2017qfl} 
  J.~Greensite and R.~Hollwieser,
  Phys.\ Rev.\ D {\bf 97}, no. 11, 114504 (2018)
  [arXiv:1708.08031 [hep-lat]].


\bibitem{rw}
A.~Roberge and N.~Weiss, Nucl. Phys. B {\bf 275}, 734 (1986).


\bibitem{ZN_1} 
  H.~Kouno, Y.~Sakai, T.~Makiyama, K.~Tokunaga, T.~Sasaki and M.~Yahiro,
  J.\ Phys.\ G {\bf 39}, 085010 (2012).

\bibitem{ZN_2}
Y.~Sakai, H.~Kouno, T.~Sasaki and M.~Yahiro,
  Phys.\ Lett.\ B {\bf 718}, 130 (2012)
  [arXiv:1204.0228 [hep-ph]].


\bibitem{ZN_3}
H.~Kouno, T.~Makiyama, T.~Sasaki, Y.~Sakai and M.~Yahiro,
  J.\ Phys.\ G {\bf 40}, 095003 (2013)
  [arXiv:1301.4013 [hep-ph]].

\bibitem{ZN_4} 
  T.~Iritani, E.~Itou and T.~Misumi,
  JHEP {\bf 1511}, 159 (2015)
  [arXiv:1508.07132 [hep-lat]].

\bibitem{ZN_5} 
  H.~Kouno, K.~Kashiwa, J.~Takahashi, T.~Misumi and M.~Yahiro,
  Phys.\ Rev.\ D {\bf 93}, no. 5, 056009 (2016)
  [arXiv:1504.07585 [hep-ph]].



\bibitem{ZN_6} 
  A.~Cherman, S.~Sen, M.~Unsal, M.~L.~Wagman and L.~G.~Yaffe,
  Phys.\ Rev.\ Lett.\  {\bf 119}, no. 22, 222001 (2017)
  [arXiv:1706.05385 [hep-th]].

\bibitem{ZN_7} 
  Y.~Tanizaki, Y.~Kikuchi, T.~Misumi and N.~Sakai,
  Phys.\ Rev.\ D {\bf 97}, no. 5, 054012 (2018)
  [arXiv:1711.10487 [hep-th]].




\bibitem{FMRW}
M.~D'Elia and F.~Sanfilippo,
  Phys.\ Rev.\  D {\bf 80}, 111501 (2009)
  [arXiv:0909.0254 [hep-lat]];

\bibitem{OPRW}
P.~de Forcrand and O.~Philipsen,
  Phys.\ Rev.\ Lett.\  {\bf 105}, 152001 (2010)
  [arXiv:1004.3144 [hep-lat]].

\bibitem{CGFMRW}
C.~Bonati, G.~Cossu, M.~D'Elia and F.~Sanfilippo,
  Phys.\ Rev.\ D {\bf 83}, 054505 (2011)
  [arXiv:1011.4515 [hep-lat]].


\bibitem{wumeng}
L.-K.~Wu and X.-F.~Meng,
  Phys.\ Rev.\ D {\bf 87}, 094508 (2013)
  [arXiv:1303.0336 [hep-lat]].

\bibitem{PP_wilson} 
O.~Philipsen and C.~Pinke,
  Phys.\ Rev.\ D {\bf 89}, 094504 (2014)
  [arXiv:1402.0838 [hep-lat]].

\bibitem{wumeng2}
L.-K.~Wu and X.-F.~Meng,
  Phys.\ Rev.\ D {\bf 90}, 094506 (2014)
  [arXiv:1405.2425 [hep-lat]].

\bibitem{nagata15} 
  K.~Nagata, K.~Kashiwa, A.~Nakamura and S.~M.~Nishigaki,
  Phys.\ Rev.\ D {\bf 91}, 094507 (2015)
  [arXiv:1410.0783 [hep-lat]].

\bibitem{kashiwa}
K.~Kashiwa and A.~Ohnishi,
  Phys.\ Rev.\ D {\bf 93}, no. 11, 116002 (2016)
  [arXiv:1602.06037 [hep-ph]];
  Phys.\ Lett.\ B {\bf 750}, 282 (2015)
  [arXiv:1505.06799 [hep-ph]].

\bibitem{rw_physicalpoint}
C.~Bonati, M.~D'Elia, M.~Mariti, M.~Mesiti, F.~Negro and F.~Sanfilippo,
  Phys.\ Rev.\ D {\bf 93}, no. 7, 074504 (2016)
  [arXiv:1602.01426 [hep-lat]].



\bibitem{makiyama16} 
  T.~Makiyama {\it et al.},
  Phys.\ Rev.\ D {\bf 93}, 014505 (2016)
  [arXiv:1502.06191 [hep-lat]].

\bibitem{nf2PP} 
  C.~Pinke and O.~Philipsen,
  arXiv:1508.07725 [hep-lat].

\bibitem{cuteri}
C.~Czaban, F.~Cuteri, O.~Philipsen, C.~Pinke and A.~Sciarra,
  Phys.\ Rev.\ D {\bf 93}, no. 5, 054507 (2016)
  [arXiv:1512.07180 [hep-lat]].




\bibitem{model-rw}
H.~Kouno, Y.~Sakai, K.~Kashiwa and M.~Yahiro,
  J.\ Phys.\ G {\bf 36}, 115010 (2009) 
  [arXiv:0904.0925 [hep-ph]].

\bibitem{Sakai:2009dv} 
Y.~Sakai, K.~Kashiwa, H.~Kouno, M.~Matsuzaki and M.~Yahiro,
  Phys.\ Rev.\ D {\bf 79}, 096001 (2009) 
  [arXiv:0902.0487 [hep-ph]].

\bibitem{sakai2}
Y.~Sakai, T.~Sasaki, H.~Kouno and M.~Yahiro,
  Phys.\ Rev.\ D {\bf 82}, 076003 (2010) 
  [arXiv:1006.3648 [hep-ph]].

\bibitem{sakai3}
T.~Sasaki, Y.~Sakai, H.~Kouno and M.~Yahiro,
  Phys.\ Rev.\ D {\bf 84}, 091901 (2011) 
  [arXiv:1105.3959 [hep-ph]].

\bibitem{sakai4}
H.~Kouno, M.~Kishikawa, T.~Sasaki, Y.~Sakai and M.~Yahiro,
  Phys.\ Rev.\ D {\bf 85}, 016001 (2012) 
  [arXiv:1110.5187 [hep-ph]].

\bibitem{holorw}
G.~Aarts, S.~P.~Kumar and J.~Rafferty,
  JHEP {\bf 1007}, 056 (2010)  
  [arXiv:1005.2947 [hep-th]].

\bibitem{holorw2}
J.~Rafferty,
  JHEP {\bf 1109}, 087 (2011)  
  [arXiv:1103.2315 [hep-th]].

\bibitem{morita} 
K.~Morita, V.~Skokov, B.~Friman and K.~Redlich,
  Phys.\ Rev.\ D {\bf 84}, 076009 (2011) 
  [arXiv:1107.2273 [hep-ph]].

\bibitem{weise} 
K.~Kashiwa, T.~Hell and W.~Weise,
  Phys.\ Rev.\ D {\bf 84}, 056010 (2011) 
  [arXiv:1106.5025 [hep-ph]].

\bibitem{pagura} 
V.~Pagura, D.~Gomez Dumm and N.~N.~Scoccola,
  Phys.\ Lett.\ B {\bf 707}, 76 (2012) 
  [arXiv:1105.1739 [hep-ph]].

\bibitem{buballa} 
D.~Scheffler, M.~Buballa and J.~Wambach,
  Acta Phys.\ Polon.\ Supp.\  {\bf 5}, 971 (2012)
  [arXiv:1111.3839 [hep-ph]].

\bibitem{kp13}
K.~Kashiwa and R.~D.~Pisarski,
  Phys.\ Rev.\ D {\bf 87}, 096009 (2013) 
  [arXiv:1301.5344 [hep-ph]].

\bibitem{rw-2color} 
K.~Kashiwa, T.~Sasaki, H.~Kouno and M.~Yahiro,
  Phys.\ Rev.\ D {\bf 87}, 016015 (2013)  
  [arXiv:1208.2283 [hep-ph]].

\bibitem{nf3_1st} 
  P.~de Forcrand and M.~D'Elia,
  PoS LATTICE {\bf 2016}, 081 (2017)
  [arXiv:1702.00330 [hep-lat]].

\bibitem{piswil}
R.~D.~Pisarski and F.~Wilczek,
  Phys.\ Rev.\ D {\bf 29}, 338 (1984).


\bibitem{Bernard:2006vv} 
  C.~Bernard, M.~Golterman, Y.~Shamir and S.~R.~Sharpe,
  Phys.\ Lett.\ B {\bf 649}, 235 (2007)
  [hep-lat/0603027].

\bibitem{adjoint1} 
  F.~Karsch and M.~Lutgemeier,
  Nucl.\ Phys.\ B {\bf 550}, 449 (1999)
  [hep-lat/9812023].


\bibitem{weisz} 
P.~Weisz,
  Nucl.\ Phys.\ B {\bf 212}, 1 (1983).

\bibitem{curci} 
G.~Curci, P.~Menotti and G.~Paffuti,
  Phys.\ Lett.\ B {\bf 130}, 205 (1983)
  [Erratum-ibid.\ B {\bf 135}, 516 (1984)].

\bibitem{morning}
  C.~Morningstar and M.~J.~Peardon,
  Phys.\ Rev.\ D {\bf 69}, 054501 (2004)
  [hep-lat/0311018]. 

\bibitem{FS1}
A.~M.~Ferrenberg and R.~H.~Swendsen
  Phys.\ Rev.\ Lett. {\bf 61}, 2635 (1988);
Phys.\ Rev.\ Lett. {\bf 63}, 1195 (1989).




\bibitem{lattsp1}
  Y.~Aoki, S.~Borsanyi, S.~Durr, Z.~Fodor, S.~D.~Katz, S.~Krieg and K.~K.~Szabo,
  JHEP {\bf 0906}, 088 (2009)
  [arXiv:0903.4155 [hep-lat]]. 

\bibitem{lattsp2}
  S.~Borsanyi, G.~Endrodi, Z.~Fodor, A.~Jakovac, S.~D.~Katz, S.~Krieg, C.~Ratti and K.~K.~Szabo,
  JHEP {\bf 1011}, 077 (2010)
  [arXiv:1007.2580 [hep-lat]]; 

\bibitem{lattsp3}
S.~Borsanyi, Z.~Fodor, C.~Hoelbling, S.~D.~Katz, S.~Krieg and K.~K.~Szabo,
  Phys.\ Lett.\ B {\bf 730}, 99 (2014)
  [arXiv:1309.5258 [hep-lat]].


\bibitem{wflow_1} 
  M.~L\"uscher,
  JHEP {\bf 1008}, 071 (2010)
  Erratum: [JHEP {\bf 1403}, 092 (2014)]
  [arXiv:1006.4518 [hep-lat]].



\bibitem{wflow_2} 
  S.~Borsanyi {\it et al.},
  JHEP {\bf 1209}, 010 (2012)
  [arXiv:1203.4469 [hep-lat]].


\bibitem{Golterman:1984dn}
  M.~F.~L.~Golterman and J.~Smit,
  Nucl.\ Phys.\ B {\bf 255}, 328 (1985).

\bibitem{Kilcup:1986dg}
  G.~W.~Kilcup and S.~R.~Sharpe,
  Nucl.\ Phys.\ B {\bf 283}, 493 (1987).



\bibitem{gpu2}
C.~Bonati {\it et al.},
  Int.\ J.\ Mod.\ Phys.\ C {\bf 28}, no. 05, 1750063 (2017)
  [arXiv:1701.00426 [hep-lat]].

\bibitem{gpu3}
C.~Bonati {\it et al.},
  Int.\ J.\ Mod.\ Phys.\ C {\bf 29}, no. 01, 1850010 (2018)
  [arXiv:1801.01473 [hep-lat]].

\bibitem{gpu1} 
  C.~Bonati, G.~Cossu, M.~D'Elia and P.~Incardona,
  Comput.\ Phys.\ Commun.\  {\bf 183}, 853 (2012)
  [arXiv:1106.5673 [hep-lat]].




\bibitem{LawSarb}
  I.~D.~Lawrie and S.~Sarbach, \emph{Theory of Tricritical Points}, in C.~Domb, J.~L.~Lebowitz (eds.)
  ``Phase transitions and critical phenomena, vol. 11'', Academic Press (1987).


\bibitem{pv_rev}
A.~Pelissetto and E.~Vicari, 
  Phys. Rep. {\bf 368}, 549 (2002)
  [arXiv:cond-mat/0012164].

\bibitem{isingcrit}
H.~W.~J.~Bl\"{o}te, E.~Luijten and J.~R.~Heringa,
   J.\ Phys.\ A: Math.\ Gen.\ \textbf{28} 6289 (1995)
   [arXiv:cond-mat/9509016].


\bibitem{potts3d}
C.~Bonati and M.~D'Elia,
  Phys.\ Rev.\ D {\bf 82}, 114515 (2010).
  [arXiv:1010.3639 [hep-lat]].


\bibitem{Pelissetto:2017pxb} 
  A.~Pelissetto, A.~Tripodo and E.~Vicari,
  Phys.\ Rev.\ E {\bf 97}, no. 1, 012123 (2018)
  [arXiv:1711.04567 [cond-mat.stat-mech]].

\bibitem{Pelissetto:2017sfd} 
  A.~Pelissetto, A.~Tripodo and E.~Vicari,
  Phys.\ Rev.\ D {\bf 96}, no. 3, 034505 (2017)
  [arXiv:1706.04365 [hep-lat]].






\bibitem{Cheng:2007jq}
  M.~Cheng, N.~H.~Christ, S.~Datta, J.~van der Heide, C.~Jung, F.~Karsch, O.~Kaczmarek and E.~Laermann {\it et al.},
  Phys.\ Rev.\ D {\bf 77}, 014511 (2008)
  [arXiv:0710.0354 [hep-lat]]. 




\bibitem{critexpo2}
  M.~Campostrini, M.~Hasenbusch, A.~Pelissetto, P.~Rossi and E.~Vicari,
  Phys.\ Rev.\ B {\bf 63}, 214503 (2001)
  [cond-mat/0010360].


\bibitem{nf2paper}
  M.~D'Elia, A.~Di Giacomo and C.~Pica,
  Phys.\ Rev.\ D {\bf 72}, 114510 (2005)
  [hep-lat/0503030].

\bibitem{Bernard:1999fv} 
  C.~W.~Bernard, C.~E.~Detar, S.~A.~Gottlieb, U.~M.~Heller, J.~Hetrick, K.~Rummukainen, R.~L.~Sugar and D.~Toussaint,
  Phys.\ Rev.\ D {\bf 61}, 054503 (2000)
  [hep-lat/9908008].

\bibitem{tchot}
A.~Bazavov, T.~Bhattacharya, M.~Cheng, C.~DeTar, H.~T.~Ding, S.~Gottlieb,
R.~Gupta and P.~Hegde {\it et al.},
  Phys.\ Rev.\ D {\bf 85}, 054503 (2012) [arXiv:1111.1710 [hep-lat]].


\end{thebibliography}
\end{document}